# The Voice of Equity: A Systematic Evaluation of Bias Mitigation Techniques for Speech-Based Cognitive Impairment Detection Across Architectures and Demographics

Yasaman Haghbin, MS[1]; Sina Rashidi, MS[1]; Ali Zolnour, MS[1]; Maryam Zolnoori, PhD[1,2]
[1]Columbia University Irving Medical Center, *New York, NY 10032*
[2]*Columbia University School of Nursing, New York, NY 10032*

**Corresponding Author**
Maryam Zolnoori, PhD
Columbia University Medical Center,
School of Nursing Columbia University
Phone: 317-515-1950
**Address**: 560 W 168th St, New York, NY 10032
E-Mail: **mz2825**@cumc.columbia.edu; m.zolnoori@gmail.com

# Abstract

Speech-based detection of cognitive impairment offers a scalable, non-invasive screening approach, yet algorithmic bias across demographic and linguistic subgroups remains a critical, underexplored challenge. Addressing this gap is essential to prevent the amplification of healthcare disparities as speech-based screening tools are increasingly deployed in clinical settings.

We present the first comprehensive fairness analysis framework for speech-based multi-class cognitive impairment detection, systematically evaluating bias mitigation across model architectures, demographic subgroups, and languages. We developed and compared two distinct transformer-based architectures, SpeechCARE-AGF and Whisper-LWF-LoRA, on the multilingual NIA PREPARE Challenge dataset (N=2,058; English, Spanish, Mandarin). Unlike prior work that typically examines single mitigation techniques or limited subgroups, we conducted a comparative analysis of pre-processing (non-generative and generative oversampling including voice conversion and text-to-speech synthesis), in-processing (frequency reweighting, dynamic TPR reweighting, adversarial debiasing, and language-specific representation enhancement), and post-processing (calibrated equalized odds) approaches. We assessed baseline fairness and mitigation effectiveness using Equality of Opportunity and Equalized Odds across multiple demographic variables: gender, age, education, and languages.

Both models achieved strong baseline performance (SpeechCARE-AGF: F1=70.87±0.21, Whisper-LWF-LoRA: F1=71.46±0.34), but equality of opportunity and equalized odds assessment revealed substantial fairness disparities across demographic subgroups. Adults aged ≥80 showed lower true positive rates (TPR: ~46%) compared to younger groups (~77-83%), indicating significant violations of equality of opportunity by age. Spanish speakers demonstrated reduced sensitivity (TPR: 44-48%) relative to English speakers (~57-62%). Our systematic comparison revealed that mitigation effectiveness varied critically by architecture: oversampling improved SpeechCARE-AGF performance for older adults (80+ TPR: 46.19%→49.97%), but yielded minimal gains for Whisper-LWF-LoRA. Frequency reweighting provided the most consistent improvements across both architectures, substantially reducing age-related disparities (improving equality of opportunity: 80+ TPR: SpeechCARE-AGF 59.38%, Whisper-LWF-LoRA 55.14%) while maintaining performance in other subgroups. Combined mitigation techniques (voice conversion + frequency reweighting for SpeechCARE-AGF) and language-based frequency reweighting for Whisper-LWF-LoRA achieved the strongest reductions in equality of opportunity and equalized odds violations across subgroups.

This study addresses a critical gap in healthcare AI by providing the first comprehensive evaluation of fairness-oriented mitigation techniques for speech-based cognitive impairment detection. Through systematic comparison across architectures, demographic subgroups, and languages, we demonstrate that architectural design fundamentally shapes bias patterns and mitigation effectiveness. Adaptive fusion mechanisms enable more flexible responses to data-level interventions, while frequency-based reweighting offers robust improvements across diverse architectures. Our findings establish that fairness interventions must be tailored to both model architecture and demographic characteristics, providing a systematic framework for developing equitable speech-based screening tools essential for reducing diagnostic disparities in cognitive healthcare.

# 1. INTRODUCTION

Cognitive impairment, including mild cognitive impairment (MCI) and Alzheimer's disease (AD), affect approximately 7.2 million Americans aged 65 and older, a number projected to rise to 13.8 million by 2060.[1–3] Despite nationwide efforts, more than half of individuals with MCI and early-stage AD, go undetected in routine care, delaying timely care and increasing healthcare burden.[4–6] Early and accessible detection methods are therefore a public-health priority, particularly from the National institute on aging (NIA).[7]

Researchers have increasingly investigated speech as a non-invasive and scalable biomarker for early cognitive impairment detection. Even brief, conversational samples encode neurodegenerative changes. Neurodegenerative processes associated with cognitive impairment can affect speech production and control, resulting in altered acoustic features such as pitch, jitter, and vocal tone.[8] Concurrent memory–language decline degrades syntactic complexity, information density, and lexical diversity, often leading to disorganized and less fluent speech.[9,10] These changes have been shown to discriminate prodromal AD several years before clinical diagnosis.[11]

The integration of machine learning (ML) into speech-based cognitive impairment detection has shown considerable promise. Classical acoustic feature-engineering methods (e.g., MFCCs, prosodic statistics) and self-supervised acoustic transformer encoders (e.g., wav2vec 2.0[12], HuBERT[13]) effectively capture acoustic and temporal cues in patient speech, while linguistic transformer models (e.g., BERT[14] and its derivatives) show strong performance in detecting discourse-level anomalies such as disfluencies on benchmark transcribed speech datasets (e.g., Dementia Bank Pitt Corpus[15]). Notably, multimodal fusion—integrating both acoustic and linguistic modalities—further improves classification accuracy by modeling complementary signals.[16]

Despite advances in ML-based clinical decision support systems (CDSS), algorithmic bias remains a significant and underexplored challenge, particularly in cognitive impairment screening studies.[17] Models trained on imbalanced or unrepresentative datasets can produce systematically different results for specific demographic groups, leading to disparities in predictive performance.[18] These differences are often observed across gender, education level, and age group. For instance, Wang et al.[19] evaluated ML models for Alzheimer's progression using magnetic resonance imaging and found significantly lower predictive area-under-the-curve (AUC) for male compared to female participants (female AUC=0.56, male AUC=0.44). They also observed age-related disparities, with individuals over 80 years achieving lower predictive performance than younger groups. These disparities highlight the risks of amplifying healthcare inequities through the deployment of biased algorithms.

In ML research more broadly, bias is commonly assessed using fairness metrics such as equality of opportunity (equal true positive rates across groups) and equalized odds (balanced true and false positive rates).[18,20,21] Mitigation techniques can be applied at three stages: pre-processing, in-processing, and post-processing.[22] Pre-processing methods address bias in training data through techniques such as oversampling[22]. In-processing approaches integrate fairness constraints during model training, for example through reweighting[23] or adversarial debiasing[24]. Post-processing methods adjust model outputs to improve group-level parity.[22] Although well established in general ML applications, these strategies have rarely been evaluated in speech-based cognitive impairment detection, leaving a critical gap in the development of fair and reliable screening tools.

To address this gap, we developed and evaluated two distinct speech-based cognitive impairment detection architectures. The first model, SpeechCARE–adaptive gating fusion (SpeechCARE-AGF)[25], integrates a multilingual acoustic transformer with a multilingual transformer language model. SpeechCARE-AGF fuses these representations through an adaptive gating mechanism that assigns data-driven weights to each modality, enabling the model to emphasize acoustic or linguistic cues based on their diagnostic value. This model received a special recognition award from the NIA PREPARE Challenge Phase 2. The second, later-developed model with higher performance adapts the Whisper-medium[26] architecture, a multilingual encoder–decoder transformer trained on 680,000 hours of speech, using Low-Rank Adaptation (LoRA)[27] and a layer-weighted fusion mechanism to increase sensitivity to subtle cognitive impairment–related speech changes. Evaluating both architectures allows us to examine how model design and representation learning affect downstream fairness. Evaluating both models in parallel enables us to investigate how differences in architecture, fusion strategy, and representation learning influence predictive performance, and shape the models' responses to bias mitigation techniques.

We systematically analyze the impact of bias-mitigation techniques on both architectures using multilingual, multitask NIA PREPARE Challenge benchmark data spanning three classes (control, MCI, AD). Specifically, we assess pre-processing, in-processing, and post-processing methods and quantify their effects on subgroup bias. Unlike prior work in cognitive impairment detection, which typically evaluates bias or tests a single mitigation technique, our study provides the first comprehensive comparison of multiple mitigation techniques across two speech-based architectures in a multi-class setting, establishing a framework for systematic fairness analysis and mitigation.

## 2. METHOD

We summarized our full experimental workflow in **Figure 1**, which outlines the end-to-end fairness evaluation pipeline. The figure illustrates how speech inputs pass through our two cognitive impairment detection models, SpeechCARE-AGF and Whisper-LWF-LoRA, followed by subgroup-level fairness assessment and the application of pre-, in-, and post-processing bias-mitigation techniques.

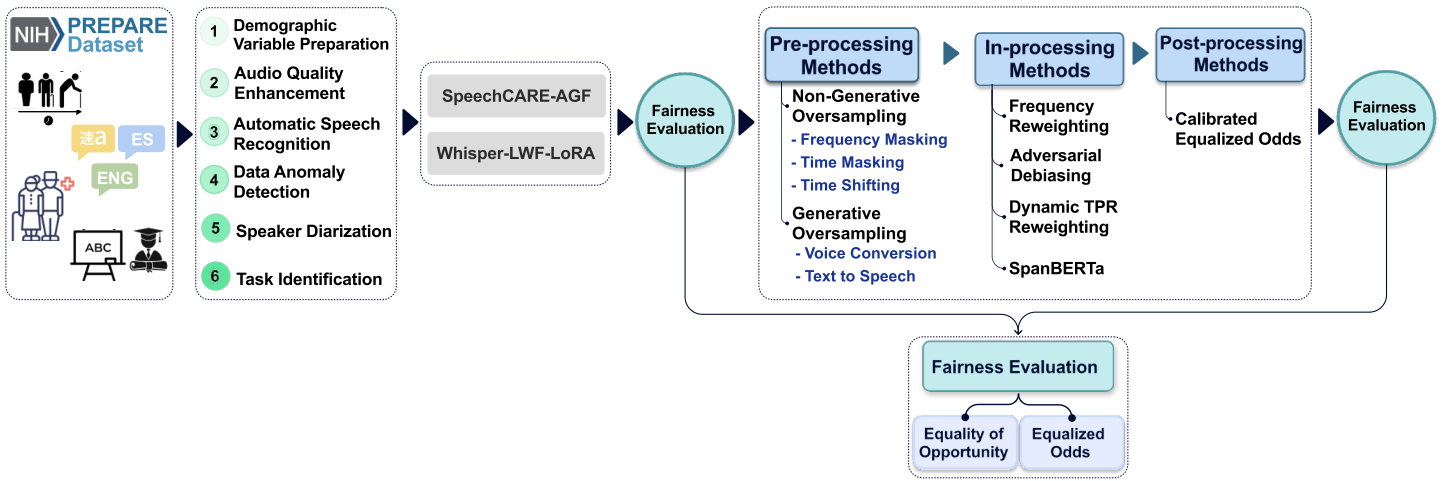

**Figure 1.** Overview of our fairness evaluation pipeline. The pipeline begins with preprocessing. After that, speech data are passed through two multimodal models. Model outputs are then subjected to fairness evaluation across demographic subgroups. Based on observed disparities, three stages of bias mitigation are applied: pre-processing, in-processing, and post-processing adjustments. The final fairness assessment quantifies improvements using Equality of Opportunity and Equalized Odds metrics.

## 2.1 Data Description

The NIA PREPARE Challenge dataset comprises speech recordings from 2,058 participants, officially divided into a train set =1,646 and test set =412). Data were drawn from 10 corpora and span three languages: English (~80%), Spanish (~17.5%), and Mandarin (~2%). Clinical diagnoses included 1,140 cognitively healthy, 268 MCI, and 650 AD. Participants were 59.8% female, aged 46–99 years. Education is reported heterogeneously, either as years of schooling or as ordinal categories (e.g., elementary, undergraduate), and is missing for ~35% of participants; race was only recorded for about 8%. Most audio recordings (66.5%) were exactly 30 seconds, followed by 20.7% between 20–30 seconds, 11.9% between 10–20 seconds, and 0.9% under 10 seconds. The distribution was similar across diagnostic classes. According to the challenge guidelines, all recordings were truncated to a maximum of 30 seconds, but the specific truncation procedure was not described. Very short recordings were likely due to issues such as incomplete or corrupted audio segments or technical failures. Importantly, the dataset contains a systematic language–label confound: All Mandarin recordings are labeled MCI and include no Mandarin control or AD samples. This imbalance prevents meaningful evaluation of model fairness for Mandarin speakers and can cause models to assign the MCI label for all Mandarin inputs. To address this limitation, we conducted an external generalizability evaluation on a balanced set of Mandarin control and MCI samples (see Section 2.7). **Figure 2** provides information about the distribution of diagnostic classes across demographic and linguistic groups.

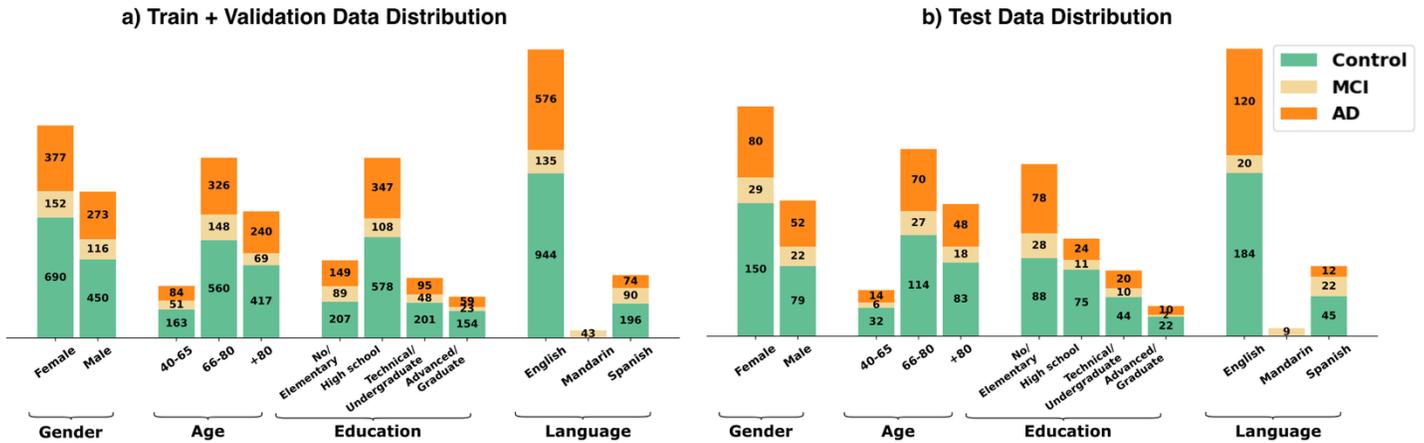

**Figure 2.** Demographic distribution of the NIA PREPARE dataset across gender, age, education, and language, separated by diagnostic categories (Control, MCI, AD).

We created a validation set by randomly selecting 20% of the training data (329 participants), while maintaining a stratified split across diagnostic classes (Control, MCI, AD) and demographic variables (age, gender, education). This ensures fair performance evaluation and prevent overfitting. This approach preserved the overall distribution of key characteristics present in the original training set. Throughout the training process for all experiments, we used this validation set for early stopping and hyperparameter tuning to avoid data leakage from the final test set.

## 2.2 Data Preprocessing

**Demographic Variable Preparation.** We used four demographic variables for bias assessment and mitigation: gender, age, language, and education. Gender (female/male) and language (English, Spanish, Mandarin) are used as recorded. Age was grouped into three categories commonly used in gerontological research: 46–65 (12%), 66–80 (52.7%), and ≥81 (35.3%).[28,29] Education was mapped to four levels aligned with the International Standard Classification of Education (ISCED-11): elementary, high school, undergraduate, and graduate, reflecting variation in cognitive reserve associated with dementia risk.[30] Because education was missing for 35.23% of participants, we used the Iterative Imputer method[31] (available in the Scikit-learn package), which models each incomplete variable as a function of other features and iteratively estimates missing values. Using age, gender, and language as predictors, this approach helped preserve the integrity of the education data by leveraging relationships among observed variables.

**Audio Quality Enhancement.** We applied an 8 kHz low-pass filter to reduce high-frequency noise while preserving essential speech components.[32] We avoided deep-learning denoising because it may distort frequency and energy features critical for cognitive-impairment detection.[25]

**Automatic Speech Recognition (ASR).** All audio was transcribed with Whisper-Large.[26] To assess transcription quality, we computed word error rates (WER)[33] on a stratified random sample: 5% of English (N=83), 10% of Spanish (N=36), and 15% of Mandarin (N=9). WERs are 0.13 (English), 0.35 (Spanish), and 0.86 (Mandarin), indicating strong English performance and moderate limitations in Spanish/Mandarin, likely reflecting phonological/grammatical differences and less availability of Spanish/Mandarin speech corpora for Whisper training.

**Data Anomaly Detection.** During transcript review, we observed that some recordings contained only clinician speech. To detect these automatically, we used LLaMA-3-405B[34] with prompt engineering. We first designed an initial prompt and evaluated it on a set of 50 transcripts containing both participant and clinician speech, along with 7 transcripts known to include only clinician speech. Based on misclassifications, we revised the prompt until it consistently differentiated participant speech from clinician-only speech. The final prompt, validated on an independent set of 50 mixed (patient and clinician) transcripts and 5 clinician-only transcripts, achieved 100% accuracy. Applied to the full dataset, it flagged 22 clinician-only transcripts, which we removed to preserve training integrity and avoid bias.

**Speaker diarization.** Most existing diarization pipelines, such as WhisperX[35] and NVIDIA NeMo[36], rely on acoustic clustering to separate speakers and assign non-identifying speaker labels (e.g., speaker0, speaker1). While effective for role-agnostic diarization, these approaches cannot provide meaningful role labels (e.g., patient vs. clinician) without an additional mapping step. This limitation makes them less suitable for clinical tasks (e.g., cognitive impairment) where patient-only speech is often required for downstream analysis.

To address this gap, we developed a two-step pipeline that combines WhisperX with LLaMA-3.1-405B for integrated diarization and role attribution. WhisperX first generated transcripts with word-level timestamps. LLaMA-405B then reconstructed sentence boundaries and performed text-based diarization, assigning each sentence to Speaker 1 or Speaker 2. It then classified each speaker as patient or clinician using linguistic cues and conversational context.

We built and evaluated this pipeline using the manually transcribed and diarized Hammond corpus of clinician–patient communication (N=200) for the picture-description task in the Pitt DementiaBank. Half of the data was used for prompt development, and the remainder for evaluation. One-shot prompting yielded the highest performance, with a Word Diarization Error Rate (WDER) of 0.09 and a speaker-count MAE of 0.10 (lower is better). This surpassed the state-of-the-art NVIDIA NeMo model (WDER=0.15; SpkCnt-MAE=0.73) on the same test set, likely due to LLMs' stronger use of linguistic structure. We tested several LLMs (LLaMA-3-8B[34], LLaMA-3-70B[34], and LLaMA-405B), and LLaMA-405B performed best.

Applied to the PREPARE dataset, this pipeline enabled extraction of patient-only audio segments (excluding clinician speech) for building the speech-processing algorithm in the next phase.

**Speech task Identification.** The PREPARE dataset features speech samples derived from multiple speech tasks (e.g., picture description, reading, story recall), but no metadata was provided to indicate which task each sample corresponds to. To classify the task type of each sample, we used LLaMA-3.1-70b-instruct with few-shot learning. To automatically infer task type, we employed LLaMA-3.1-70B-Instruct in a few-shot classification setting. Using this approach, each recording was assigned to one of six task categories: personal narrative, story recall, picture description, semantic verbal fluency, sentence reading, or voice-assistant interactions. See the details of task identification in this paper.[25]

## 2.3 Model Architecture

We propose two models for automatic cognitive impairment (MCI, AD) detection from speech: SpeechCARE-AGF and Whisper-LWF-LoRA.

## Model 1: SpeechCARE-AGF

We used SpeechCARE Adaptive Gating Fusion (SpeechCARE-AGF), a novel multimodal fusion approach that dynamically weights modalities based on the most informative feature vectors for early detection of cognitive decline. As illustrated in **Figure 3**, the model consists of two components: a feature network and a fusion network.

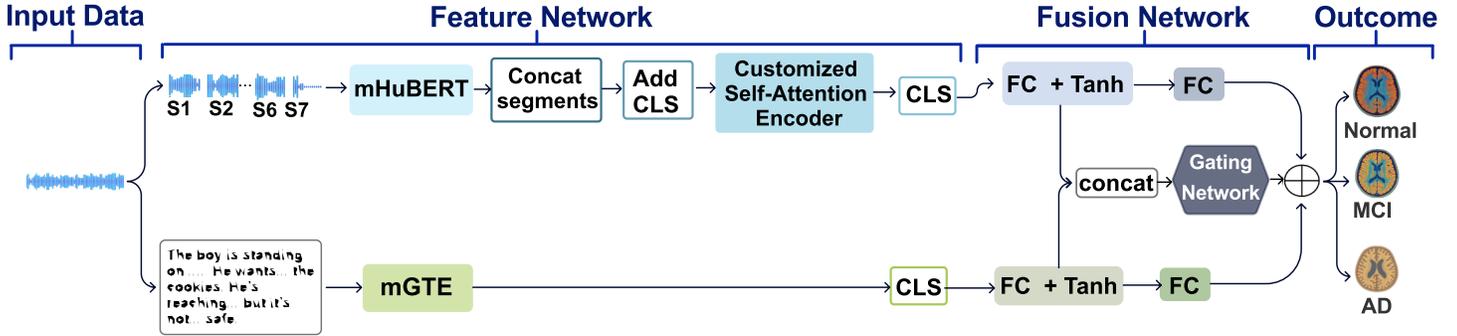

**Figure 3.** Architecture of the SpeechCARE-AGF model. The Feature Network extracts acoustic and linguistic embeddings using mHuBERT and mGTE, respectively. The acoustic branch includes a Customized Self-Attention Encoder (CSE) with two transformer blocks (four attention heads, 0.1 dropout) and a learnable [CLS] token for global context. The Fusion Network integrates both modalities through a learnable Gating Network. After each modality's [CLS] embedding passes through fully connected layers with Tanh activation, the resulting representations are concatenated and fed into a gating module composed of a linear layer followed by a Softmax activation. This module dynamically assigns weights to each modality, and the weighted modality scores are combined to produce the final prediction.

**Feature Network.** Feature network generates representations from pre-trained linguistic and acoustic transformers:

- **Linguistic branch.** Transcripts were processed using mGTE[37], a multilingual encoder-only transformer with 305 million parameters trained on more than 1 trillion tokens across 75 languages. Compared with models such as BERT[38], mGTE supports substantially longer context windows (up to 8,192 tokens), enabling more effective modeling of discourse-level structure, including long-range dependencies, disfluencies, and syntactic irregularities that are commonly observed in speech associated with cognitive decline. The model was fine-tuned using standard transformer fine-tuning procedures. Given a tokenized sequence $L = \{t_{CLS}, t_1, ..., t_N\}$, mGTE produces contextualized embeddings $\{e_{CLS}, e_1, ..., e_N\}$. The final-layer $[CLS]$ embedding $e_{CLS}$ provides a summary linguistic representation for each transcript and served as the input to the fusion network.

- **Acoustic branch.** Raw audio was encoded using mHuBERT[39], a multilingual acoustic transformer pretrained on >90,000 hours of speech. However, like other self-supervised speech transformers, mHuBERT was pretrained on short segments (<5 seconds) and does not generalize well to long continuous recordings: 30-second inputs produce very long frame sequences, weakening attention patterns and making fine-tuning unstable.[25,40] To address these limitations, **(1)** each waveform $A$ was divided into overlapping 5-second segments, $A \rightarrow \{s_1, s_2, ..., s_M\}$, and each segment was encoded with mHuBERT to obtain frame-level embeddings, $f_{mHuBERT}(s_i) = e_1^{(i)}, ..., e_T^{(i)}$. All segment embeddings were concatenated into a single sequence, $E = \{e_1, ..., e_K\}$, $K = M \times T$. **(2)** To capture global context across segments, a learnable global $[CLS]$ token was prepended to the sequence: $\bar{E} = \{e_{CLS}, e_1, ..., e_K\}$. This sequence was passed through a customized self-attention encoder, a lightweight transformer with two stacked blocks and four attention heads. The final $[CLS]$ output $mHuBERT: e_{CLS} \rightarrow \bar{x}_A$ served as the acoustic representation.

**Fusion Network.** To integrate the acoustic and linguistic representations produced by the feature network, we developed the Adaptive Gating Fusion (AGF) module, which dynamically weights modalities rather than relying on fixed concatenation. This is important because different speech tasks and impairment profiles emphasize different cues, making static fusion strategies suboptimal.

Let $\bar{x}_L$ and $\bar{x}_A$ denote the linguistic and acoustic representations extracted from mGTE and mHuBERT, respectively. For each modality $m \in \{L, A\}$, the representation is projected into a hidden vector: $h_m = tanh(W_m \bar{x}_m + b_m)$. To determine each modality's contribution to the final prediction, the hidden vectors are concatenated and passed through a gating layer to generate modality attention weights: $\{\alpha_L, \alpha_A\} = Softmax(W_g[h_L; h_A] + b_g)$. Concurrently, each modality produces its own output score: $o_m = W_o h_m + b_o$. The final class probabilities are computed as a weighted combination of modality-specific scores:

$$\hat{y} = Softmax(\alpha_L o_L + \alpha_A o_A)$$

This adaptive weighting enables the model to emphasize the most informative modality for each sample and outperformed intermediate and late fusion baselines in our evaluations.

This architecture won the Special Recognition Prize in Phase 2 of the NIA Challenge. See details in the result section.

**Hyperparameter Tuning.** To train the SpeechCARE model, both the mHuBERT and mGTE transformer encoders were fine-tuned concurrently within a unified architecture. To better control the optimization process, we adopted a multi–learning rate strategy: one learning rate was assigned exclusively to mGTE parameters, and a separate learning rate was used for the remaining components of the model, including mHuBERT, the gating network, and fully connected layers.

A comprehensive hyperparameter search covering optimization parameters (learning rate, weight decay) and model-specific settings (hidden neurons, CSE attention heads) was conducted by training each configuration for 15 epochs and selecting the checkpoint with the highest validation F1-score for test evaluation. The optimal configuration used a learning rate of $1 \times 10^{-6}$ for mGTE and $1 \times 10^{-5}$ for other components, weight decay of $1 \times 10^{-3}$, batch size of 4, a CSE module with two blocks and four attention heads (dropout 0.1), fully connected layers with 128 neurons and Tanh activation.

## Model 2: Whisper-LWF-LoRA

We used Whisper-LWF-LoRA (Layer-Weighted Fusion with Low-Rank Adaptation), built on Whisper-medium[26], a multilingual encoder–decoder transformer pre-trained on approximately 680,000 hours of speech. As illustrated in **Figure 4**, the model comprises two components: a feature network, which extracts task-adapted speech representations from selected Whisper layers, and a fusion network, which integrates encoder- and decoder-derived features for classification.

**Feature Network.** The feature network operates by selectively adapting a subset of Whisper's encoder and decoder layers and aggregating their hidden representations into fixed-dimensional embeddings. Rather than fine-tuning the full model (24 layers), we empirically evaluated which layers contribute most effectively to downstream cognitive status classification. Specifically, we compared multiple layer-selection strategies—including adapting the top 12, top 10, top 6, and only the final layer—and observed that adapting the top six encoder layers and top six decoder layers yielded the most stable and accurate performance. Based on this result, Low-Rank Adaptation (LoRA) modules were inserted into the attention and feed-forward projections of encoder layers 18–24 and decoder layers 18–24, while all remaining parameters were frozen.

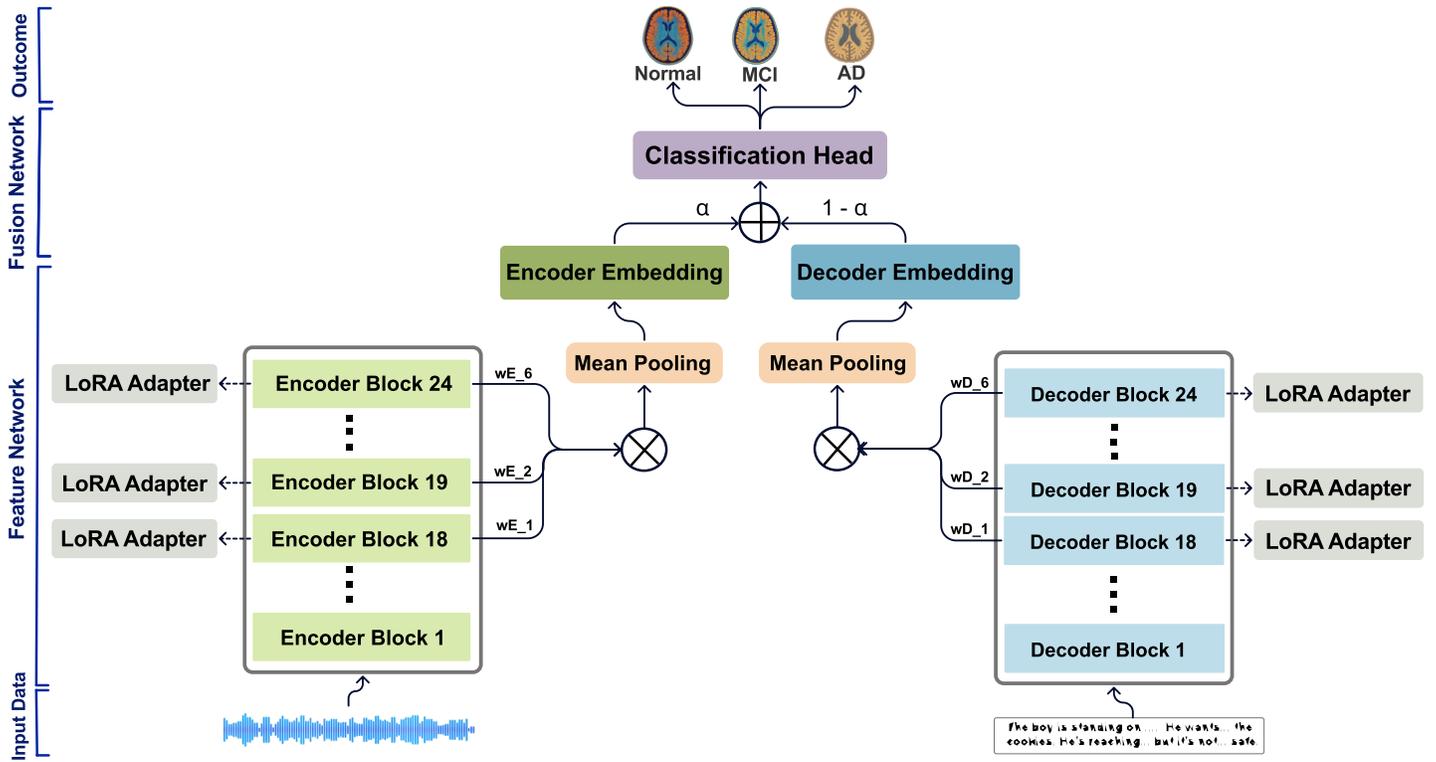

**Figure 4.** Architecture of the Whisper-LWF-LoRA model. Whisper-LWF-LoRA is a lightweight adaptation of the Whisper-medium encoder–decoder model for speech-based AD detection. The framework integrates Low-Rank Adaptation (LoRA) modules into the top six encoder and decoder layers while freezing the remaining layers. A layer-weighted fusion mechanism aggregates representations across the top transformer blocks of both encoder and decoder, producing learnable encoder and decoder embeddings. These embeddings are combined through a learnable gating coefficient (α) that adaptively balances acoustic and linguistic cues. The fused representation is then passed through a classification head to predict three diagnostic categories.

Let $E_i$ and $D_i$ denote the outputs of the $i$-th adapted encoder and decoder layers ($i = 18, …, 24$). We applied learnable layer-weighted aggregation within each component, and the weighted outputs are mean-pooled to form embeddings:

$$E_{enc} = \text{MeanPooling}(\textstyle\sum_{i=18}^{24} w_i^E . E_i), \qquad E_{dec} = \text{MeanPooling}(\textstyle\sum_{i=18}^{24} w_i^D . D_i)$$

*where $w_i^E$ and $w_i^D$ are Softmax-normalized trainable scalars.*

These yield two final embeddings: a layer-weighted encoder embedding $E_{enc}$ and a layer-weighted decoder embedding $E_{dec}$, which are used in the subsequent fusion network.

**Fusion Network.** To integrate the Whisper's encoder and decoder representations produced by the feature network, we fused them through a learnable convex interpolation: $H = \alpha E_{enc} + (1 - \alpha)E_{dec}$, where $\alpha \in [0,1]$ is a trainable gating coefficient. The fused representation $H$ is passed through a lightweight classification head composed of a Tanh-activated hidden layer $z = tanh(W_h H + b_h)$, followed by a linear projection $\hat{y} = Softmax(W_c z + b_c)$, resulting in predicted probabilities across the three diagnostic categories.

**Hyperparameter Tuning.** To train the Whisper-LWF-LoRA model, LoRA adapters were inserted into the top six encoder and decoder layers, while all other Whisper parameters remained frozen. A comprehensive hyperparameter search was conducted across optimization parameters (learning rate, weight decay, LoRA rank, LoRA dropout) and model-specific settings (hidden layer size, batch size). Each configuration was trained for 5 epochs, and the checkpoint with the highest validation F1-score was selected for test evaluation. The optimal configuration used a LoRA rank of 128, LoRA dropout of 0.05, learning rate of $1\times10^{-4}$, weight decay of $1\times10^{-3}$, batch size of 4, and 128 hidden neurons in the classification head with Tanh activation.

## 2.4 Model Evaluation

We began our evaluation by analyzing model performance. Following training, we selected the model checkpoint that achieved the highest F1-score on the validation set and used this checkpoint to evaluate performance on the official Test set released by the PREPARE challenge organizers.

To assess model stability, we repeated training five times using five distinct random seeds ($S_1$–$S_5$). Each seed controlled two sources of randomness: (1) weight initialization and (2) the order in which training batches were loaded. Running multiple seeds allowed us to quantify performance variability attributable to randomness rather than model design. For each architecture, we report the mean Test-set metrics across the five runs with 95% confidence intervals.

To assess model effectiveness, we employed several evaluation metrics. Evaluation metrics included the area under the receiver operating characteristic curve (AUC-ROC), computed using a one-vs-rest scheme across the three diagnostic classes (Control, MCI, and AD). We additionally computed F1-score, as well as AUPRC (area under the precision–recall) curves using the one-vs-rest approach.

## 2.5 Fairness Evaluation

We evaluated model bias by comparing performance subgroups defined by protected attributes $A = (gender, age, education, language)$. Let $X = \{x_1, x_2, \ldots x_i \ldots x_n\}$ denotes the inputs, $Y = \{y_1, y_2, \ldots y_i, \ldots y_n\}$ the true labels, and $\hat{Y} = h_a(X)$ the model prediction. Each class label is denoted $c \in \{control, MCI, AD\}$, and each subgroup value of attribute $A$ is denoted $a \in A$. For each class $c$ and subgroup $a$, we computed:

$$\text{True positive Rate(TPR): } TPR_{c,a} = P\big(\hat{Y} = c \mid Y = c, A = a\big)$$

$$\text{False positive Rate(FPR): } TPR_{c,a} = P\big(\hat{Y} = c \mid Y \neq c, A = a\big)$$

These correspond to the probability that the model correctly (TPR) or incorrectly (FPR) predicts class $c$ for members of subgroup $a$. Fairness was assessed using two criteria widely applied in healthcare ML:

**Equal opportunity:** requires parity in TPRs across subgroups for each class $c$:

$$TPR_{(c,a_i)} = TPR_{(c,a_j)}, \qquad \forall(ai, aj) \in A, \quad \forall c \in \{control, MCI, AD\}$$

**Equalized odds:** requires parity in both TPRs and FPRs across subgroups:

$$TPR_{(c,a_i)} = TPR_{(c,a_j)} \text{ and } FPR_{(c,a_i)} = FPR_{(c,a_j)} \qquad \forall(ai, aj) \in A, \quad \forall c \in \{control, MCI, AD\}$$

Because our task is multiclass, metrics were computed separately for each class and subgroup, then aggregated by macro-averaging:

$$MacroAvg_a(Metric) = \frac{1}{|C|} \sum_{c \in C} Metric\,(c, a),$$

where $Metric(c, a)$ is either $TPR\,(c, a)$ or $FPR\,(c, a)$ computed for class $c$ in subgroup $a$.

To assess these fairness criteria, we visualized TPR and FPR distributions across all subgroups (See Result Section). Violations of equal opportunity are indicated by disparities in TPR across subgroups, while violations of equalized odds are indicated by disparities in either TPR or FPR.

## 2.6 Bias Mitigation Techniques

To understand and reduce demographic bias in cognitive impairment detection, we systematically evaluated the effect of multiple bias-mitigation techniques across both of our speech-based architectures: SpeechCARE-AGF and Whisper-LWF-LoRA. These architectures differ in model capacity, and feature representation, making it necessary to examine how mitigation behaves under each paradigm. We assessed pre-processing, in-processing, and post-processing mitigation techniques and quantify their impact on subgroup fairness.

### Preprocessing Techniques

For pre-processing–based bias mitigation, we used both non-generative and generative oversampling techniques. Non-generative methods were chosen to introduce spectrogram-level variability without altering linguistic content. Generative methods, including text-to-speech (TTS) synthesis and voice conversion (VC), were used to create synthetic samples that mimic linguistic, prosodic, and speaker-level patterns, increasing representation of underrepresented groups beyond what can be achieved through non-generative augmentation alone.

**(a) Non-generative Oversampling.** We adapted SpecAugment, a data augmentation method introduced by Park et al.[41], which applies simple transformations to spectrograms to increase temporal and spectral diversity while preserving essential acoustic structure. In prior work, SpecAugment has been primarily used as an augmentation technique to expand training datasets and boost model performance in clinical voice analysis, including voice pathology detection[42], cough-based COVID-19 classification[43], and more recently in dementia-related speech tasks[44,45]. Building on these applications, we applied SpecAugment not as a general performance enhancer but as an oversampling strategy targeted at underrepresented or lower-performing subgroups for bias mitigation.

- **Time Shifting**[40,41] **(Figure 5.a, and 5.b).** Each audio file was circularly shifted forward or backward by a random offset. The shift was drawn from a uniform distribution between 1 second and K seconds, where K corresponded to 50% of the total audio duration.

- **Frequency Masking**[40,41] **(Figure 5.d).** Audio waveforms were converted to 128-channel log-mel spectrograms $(FFT [\text{ Fast Fourier Transform}] \ window = 2048 \ samples, \ hop \ length = 512)$. A contiguous block of $f \ mel \ channels$ with $f \sim U[1,60]$ was masked starting at a random index $f_0 \in [0, v - f)$, where $v = 128$. This controlled perturbation promotes model generalization while preserving clinically informative acoustic cues.

- **Time masking** [40,41] **(Figure 5.e).** Audio waveforms were to 128-channel log-mel spectrograms $(FFT \ window = 2048 \ samples, \ hop \ length = 512)$. A continuous block of $t$ consecutive time frames was masked, where $t \sim U[1, 60]$. The starting index $t_0$ was drawn from $[0, \tau - t)$ with $\tau$ denoting the total number of time frames. This controlled perturbation promotes robustness to temporal variability while preserving clinically informative acoustic cues.

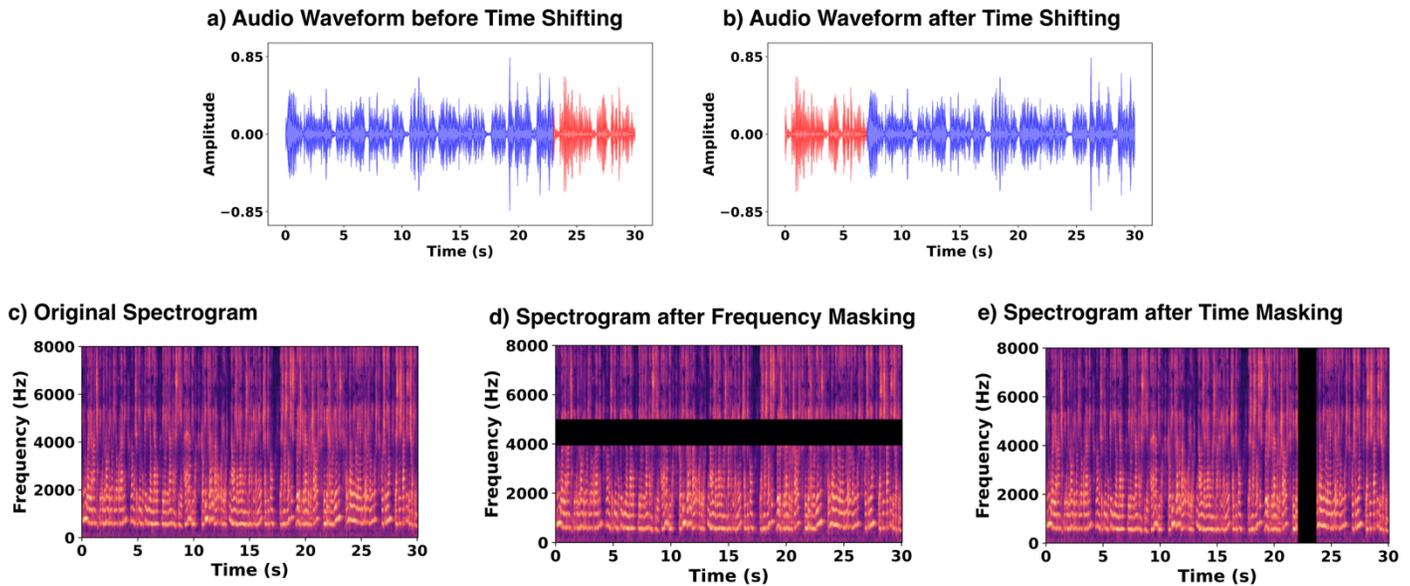

**Figure 5.** Audio data augmentation techniques applied to speech samples.

**(b) Generative Oversampling – Text to Speech.** In healthcare, text-to-speech (TTS) is widely used to support information accessibility and communication, including converting clinical text into speech and assisting patients with speech impairments (e.g., aphasia[46], esophageal speech[47]). TTS has also been incorporated into diagnostic and telemedicine systems to enable spoken interaction with clinical information. [48,49]In contrast, the use of TTS for data augmentation in speech-processing pipelines remains limited. Naeini et al.[50] demonstrated its utility by augmenting dysarthric speech with Tacotron 2[51] and Google TTS to improve automatic speech recognition performance. Beyond clinical applications, TTS-based augmentation has improved emotion recognition and spoken language identification.[52]

To our knowledge, TTS has not previously been used as an explicit oversampling strategy for bias mitigation. Most prior work synthesizes speech directly from existing transcripts. In contrast, as shown in Figure 6, our approach fine-tunes large language models (LLMs) to generate subgroup- and diagnosis-specific synthetic text, which is subsequently converted to audio using TTS, enabling demographically targeted oversampling. Building on our previous study[53] (LLMCARE framework for synthetic text generation using LLMs), this generative oversampling framework is implemented as a three-step pipeline (step1-3) comprising LLM-based synthetic text generation followed by TTS-based speech synthesis.

- **Step 1. Selection and evaluation of LLMs for synthetic language generation.** This step includes LLMs selection, prompt engineering, fine-tuning configuration and evaluation.

    **LLMs.** We fine-tuned three open-source LLMs: LLaMA-3.1-8B-Instruct, LLaMA-3-70B, and Ministral-8B-Instruct[54] These models were selected based on prior evidence of their ability to capture linguistic cues relevant to cognitive impairment (e.g., repetition, filler words). Open-source models were used in accordance with NIA PREPARE Challenge guidelines, which prohibit uploading participant data to commercial systems.

    **Prompt engineering.** Prompts incorporated the diagnostic label (Control, MCI, AD), demographic attributes (age, gender, education), and participant language. For example, when oversampling the 80+ age group and Spanish-speaking participants, the prompt explicitly indicated that the generated text should reflect speech from participants aged 80 or older and from Spanish speakers. This ensured that the generated data covered all demographic and linguistic subgroups present in the dataset.

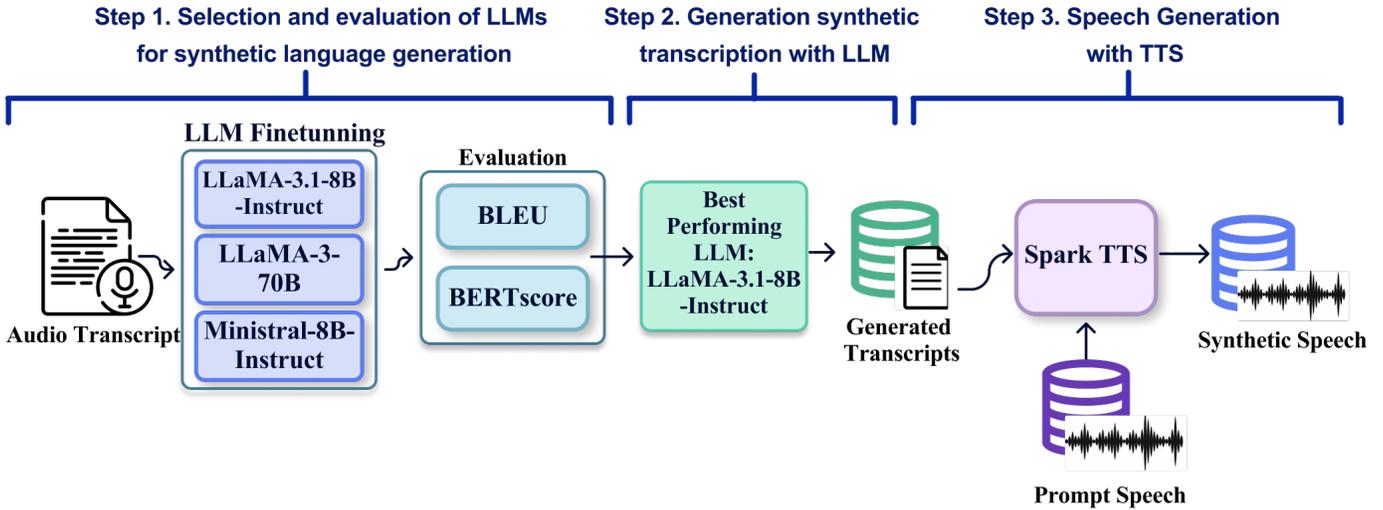

**Figure 6.** Text-to-Speech (TTS) Generative Oversampling Pipeline. Overview of the three-stage pipeline used to generate synthetic speech for augmentation. Step 1: Multiple LLMs (LLaMA-3.1-8B-Instruct, LLaMA-3-70B, Ministral-8B-Instruct) are fine-tuned and evaluated using BLEU and BERTScore to identify the best model for subgroup-specific transcript generation. Step 2: The top-performing LLM (LLaMA-3.1-8B-Instruct) generates synthetic diagnostic- and demographic-conditioned transcripts. Step 3: Synthetic transcripts are converted into waveform audio using SparkTTS, producing subgroup-targeted synthetic speech for oversampling.

**Fine-tuning configuration.** For the three LLM models, we fine-tuned models using the Quantized Low-Rank Adapter (QLoRA) framework, which inserts lightweight adapters into frozen models to enable memory-efficient training. We tested LoRA ranks of 64 and 128. Additional details are provided in our prior work.[53]

**Evaluation.** Synthetic text quality was assessed using BLEU[55] (n-gram overlap, n = 1-4) for syntactic similarity and BERTScore[56] for semantic similarity to the original transcript. Among the tested LLMs, LLaMA-3.1-8B-Instruct achieved the highest scores and was selected as the primary generator for TTS-based oversampling.

- **Step 2. Generating Synthetic Transcripts with LLM.** In this step we used LLaMA-3.1-8B-Instruct as the primary model to generate synthetic transcript.

  **Prompt template.** The prompt template is defined as: $P = f(Y, A, l, C_Y)$, where $Y$ is the diagnostic label (Control, MCI, AD), $A$ the demographic attributes (age, gender, education, language), $l$ the spoken language, and $C_Y$ the diagnosis-specific linguistic cues.

  **Synthetic text generation.** Guided our previous study[53,57], at inference, diagnosis-specific cues were excluded from prompts to avoid repetitive or unnatural outputs. Synthetic transcripts were generated from the fine-tuned LLaMA-3.1-8B model with the following inference parameters:

  $$T_j^{syn} \sim LLM_\theta(P(Y, A, l); top-p = 0.95, top-k = 50, temprature = 1), \qquad j = 1, ..., m$$
  $$\textit{where } m \textit{ is the number of synthetic transcripts generated.}$$

  We set top-p = 0.95, top-k = 50, and temperature = 1 to allow lexical and syntactic variation while maintaining discourse patterns consistent with the source clinical speech tasks.

- **Step 3. Speech Generation with TTS.** Synthetic speech was generated using Spark-TTS-0.5B, a pretrained multilingual text-to-speech model. Unlike encoder–decoder architectures such as Tacotron 2, Spark-TTS is a decoder-only transformer trained directly on speech tokens.

**Inputs.** Each synthetic transcript $T_j^{syn}$ from Step 2 was paired with a 10-second reference speaker embedding $E_s$, extracted from real audio of a participant with matching diagnostic labels. The embedding encodes speaker-specific voice characteristics, such as timbre and prosody.

**Synthetic speech generation process.** Spark-TTS autoregressively predicted speech tokens conditioned on both the transcript and the embedding, which were then converted into audio waveforms by a vocoder:

$$W_j = TTS(T_j^{syn}, E_s), \qquad j = 1, \dots, m$$

*where $W_j$ is the generated waveform and $m$ is the number of synthetic transcripts.*

To preserve label–voice alignment, embeddings $E_s$ were matched by diagnostic labels (e.g., AD transcripts synthesized with AD speaker embeddings). This process enabled us to generate synthetic samples that preserved the linguistic and demographic characteristics of the training population.

**(c) Generative Oversampling – Voice Conversion.** Voice conversion (VC) has been used for clinical applications to alter a speaker's voice while preserving the spoken content. Prior work has demonstrated its utility in several health-related contexts: Zhao et al.[58] used VC to transform the dysarthric speech of patients with amyotrophic lateral sclerosis (ALS) into clearer and more intelligible speech, and Li et al.[59] used VC to enhance the ASR performance of dysarthric-like speech in low-resource languages. Moreover, Tayebi Arasteh et al.[60] used VC for speech anonymization to increase privacy while maintaining crucial health-related information in speech. Additional studies by Qian et al.[61] and Mehrez et al.[62] leveraged VC to improve the naturalness and intelligibility of electrolarynx and dysarthric speech, respectively. VC has also been used in limited cases as a data augmentation technique. For example, Illa et al.[63] applied VC to increase the diversity of atypical speech for ASR training. However, the use of voice conversion specifically as a bias mitigation oversampling strategy, particularly in cognitive impairment detection, remains unexplored. To the best of our knowledge, our study is the first to investigate VC as a targeted oversampling method for reducing demographic and linguistic bias.

In this study, our VC pipeline consisted of two steps:

- **Step 1. Speaker-pair extraction (source → target):** We introduce a novel graph-theoretic approach to speaker pairing that maximizes acoustic diversity across the entire dataset by formulating the pairing problem as a maximum-weight matching optimization task. Each speaker in the NIA PREPARE dataset was represented using three acoustic features—VoicedSegmentsPerSec, shimmerLocaldB_sma3nz_amean, and mfcc1_sma3_stddevNorm. These features were extracted from eGeMAPS[64], a standardized set of low-level acoustic descriptors widely used in clinical and affective speech analysis for capturing physiologically meaningful aspects of voice production. We selected these specific features because, in our prior work with a reference dataset containing MMSE scores[65], they showed the strongest associations with cognitive-status measures.

Specifically, each speaker $s_i$ was represented by a normalized vector $x_i \in \mathbb{R}^3$ containing these three eGeMAPS features. We quantified acoustic dissimilarity between speakers $s_i$ and $s_j$ using cosine distance:

$$d(i,j) = 1 - \frac{xi \cdot xj}{\parallel xi \parallel_2 \parallel xj \parallel_2}$$

We constructed an undirected graph $G = (S, E)$, where speakers formed the vertices and edge weights corresponded to $d(i,j)$. To obtain diverse speaker pairs, we applied maximum-weight matching:

$$M^* = arg \, max \sum_{(i,j) \in M, \; M \subseteq E} d(i,j),$$

subject to each speaker appearing in at most one pair. This optimization-based pairing strategy represents a departure from ad-hoc selection methods typically used in voice conversion studies. Empirically, this method produced more diverse pairings and higher validation F1-scores than random pairing or simple minimum/maximum-difference pairings, demonstrating that principled speaker selection is crucial for effective bias mitigation through voice conversion.

- **Step 2. Voice conversion using FreeVC[66] (synthetic sample generation):** Each speaker pair $(s_p, s_q) \in M^*$ identified by maximum-weight matching was processed using FreeVC, a multilingual zero-shot VC model. FreeVC disentangles speech into (1) linguistic content, encoded by a pretrained prior model (WavLM), and (2) speaker identity, encoded by a speaker embedding network. This separation enables high-quality conversion across arbitrary source–target pairs without parallel data or retraining.

  For a given utterance $u_{p,r}$ ($the\ r^{th}\ recording\ from\ source\ speaker\ s_p$), FreeVC generated a converted utterance as $\tilde{u}_{p \rightarrow q,r}$: $\tilde{u}_{p \rightarrow q,r} = VC(u_{p,r}, \mathrm{E}s_q)$, where $\mathrm{E}s_q$ is the embedding of the target speaker $s_q$. The output preserved the linguistic content of $s_p$ while transforming acoustic characteristics (e.g. timbre, prosody) to resemble $s_q$.

  To maximize augmentation, each pair was processed bidirectionally: $s_p \rightarrow s_q$ and $s_q \rightarrow s_p$. Thus, if $m$ utterances were selected for VC, they formed $m/2$ disjoint source-taregt pairs and yielded $m$ converted samples, ensuring balanced coverage across both speakers.

**Oversampling Technique.** Across all pre-processing methods, we applied a consistent oversampling strategy. Specifically, we oversampled: (a) underrepresented demographic subgroups (e.g., males, ages 40–65, graduate-educated participants, Spanish speakers), and (b) subgroups showing lower fairness scores based on Equal Opportunity and Equalized Odds analyses (e.g., adults aged 80+). We also evaluated cross-demographic oversampling (e.g., female + Spanish, female + age 80+).

## In-preprocessing Techniques

For in-processing techniques, we tested four complementary strategies: (a) frequency-based reweighting, (b) adversarial debiasing to encourage demographic-invariant latent representations, (c) Dynamic True Positive Rate (TPR) reweighting, and (d) Spanish-specific representation enhancement using SpanBERTa[67] as a Spanish-specialized text encoder. These methods not only mitigated bias during training but also revealed architectural differences in how our two models respond to in-processing techniques.

**(a) Frequency-based Reweighting.** We implemented frequency-based reweighting to adjust for imbalances across subgroup–label combinations. For each training sample $i$, a weight $w(a_i, y_i)$ was computed based on the ratio of the expected to the observed frequency of its subgroup $a_i$ (as defined by the protected attributes $A$) and true class label $y_i$. This assigns higher weights to subgroup–label pairs that are underrepresented and lower weights to those that are overrepresented. The reweighted training objective was: $L = \sum_{i=1}^{n} w(a_i, y_i) l(h_a(x_i), y_i)$, where $l$ is the loss function and $h_a(x_i)$ is model prediction for sample $i$.

Reweighting was applied across all subgroups (e.g., gender, age, education, language; see Results section). The resulting weights were incorporated directly into the model's optimization procedure, ensuring that underrepresented subgroup–label pairs contributed proportionally more during training.

**(b) Adversarial Debiasing.** Speech-based cognitive-impairment detection models are prone to demographic bias because acoustic and linguistic features often encode attributes such as gender, age, and language. For example, pitch, prosody, and rhythm differ between male and female speakers, while articulation rate and lexical choice vary by age and education. These patterns can act as shortcut features, causing the model to associate demographic cues rather than cognitive markers with diagnostic labels. As a result, predictive performance may vary across subgroups, reinforcing disparities in cognitive screening. To address this, we applied adversarial debiasing, which trains the model to learn representations predictive of cognitive status while reducing sensitivity to demographic attributes.

Let $X$ denote the multimodal input (acoustic and linguistic features), $Y \in \{Control, MCI, AD\}$, and $A$ a protected attribute (e.g., gender, age, education, or language). The models (SpeechCARE-AGF, and Whisper-LWF-LoRA) encoder $E_\emptyset$ produce a shared latent representation $z = E_\emptyset(X)$, which is used by two prediction heads: a primary classifier $f_\theta(z)$ predicting $Y$, and an adversarial classifier $g_\psi(z)$ predicting $A$.

The total objective balances diagnostic accuracy and fairness:

$$\mathcal{L}_{total} = \mathcal{L}_{ADRD}(f_\theta(z), Y) - \lambda \, \mathcal{L}_{adv}(g_\psi(z), A),$$
*where both losses are cross-entropy functions, and $\lambda$ controls the strength debiasing.*

A Gradient Reversal Layer (GRL)[68] connects the encoder and the adversarial classifier, multiplying the adversarial gradient by $-\lambda$ during backpropagation. This encourages the encoder (mHuBERT and mGTE in SpeechCARE-AGF and Whisper-medium in Whisper-LWF-LoRA) to remove demographic information from $z$ while preserving features relevant to cognitive status.

In both models, we tuned the adversarial weighting coefficient ($\lambda$) to balance fairness and predictive accuracy. Values of {0.025, 0.05, 0.1, 0.2, 0.3} were tested, and the optimal $\lambda$ in each model was selected based on its ability to minimize fairness gaps while maintaining strong validation performance. In our experiments, $\lambda = 0.05$ provided the best trade-off between subgroup parity and overall F1. The adversarial debiasing framework was applied separately to each demographic attribute in each model, reducing the sensitivity of learned speech representations to subgroup-specific variations.

**(c) Dynamic True Positive Rate (TPR) Reweighting.** In-processing fairness methods such as FairBatch[69] and Group DRO (distributionally robust optimization)[70] dynamically adjust the optimization process to improve subgroup performance. FairBatch adaptively modifies minibatch composition to oversample disadvantaged subgroups, reducing disparities in Equal Opportunity and Equalized Odds. Group DRO minimizes the worst-case subgroup loss, improving robustness for underrepresented groups.

Inspired by these approaches, we introduce Dynamic True Positive Rate (TPR) Reweighting, a lightweight in-processing method designed to directly target the TPR metric during training. Instead of adjusting minibatch sampling or optimizing the worst-case loss, this method periodically updates sample weights based on each subgroup's running TPR, emphasizing subgroups where the model is underperforming.

For each demographic subgroup $a \in A$ and diagnostic class $c \in Y$, we compute the true positive rate as

$$TPR_{a,c} = \frac{Correct_{a,c}}{Total_{a,c}},$$

*where $Correct_{a,c}$ iis the number of correctly predicted samples in subgroup $a$ for class $c$, and $Total_{a,c}$ is the number of samples in that subgroup-class pair.*

Each instance $i$, belonging to subgroup $a_i$ and class $c_i$, is assigned a weight inversely proportional to its subgroup TPR: $w_i = \frac{1}{TPR_{a_i,c_i}}$. The model is trained with a weighted loss function,

$$\mathcal{L} = \frac{1}{N}\sum_{i=1}^{N} w_i \ell_i \quad \text{Where } \ell_i \text{ is the per-sample loss (e.g. cross-entropy).}$$

Low-TPR subgroups thus receive higher effective gradients, driving the network to improve sensitivity where performance lags.

**(d) Language-Specific Representation Enhancement Using SpanBERTa**. To address the pronounced performance disparities observed in Spanish-speaking participants, we incorporated SpanBERTa[67] as a language-specialized encoder for the SpeechCARE-AGF model. SpanBERTa is pretrained exclusively on large-scale Spanish corpora and captures Spanish-specific syntactic, morphological, and semantic patterns more effectively than multilingual models. The Spanish subset of the dataset (225 training, 56 validation, and 79 test samples) was used to fine-tune this encoder, and during inference, all Spanish samples were routed through the SpanBERTa encoder.

We applied this strategy only to SpeechCARE-AGF because its modular architecture naturally supports replacing the linguistic encoder on a per-language basis. In contrast, Whisper-LWF-LoRA tightly integrates linguistic and acoustic information within Whisper's decoder pathway, preventing the use of external language-specific text encoders.

### Post Processing Techniques

As a post-processing bias mitigation technique, we implemented a multi-class one-vs-rest post-processing procedure inspired by Calibrated Equalized Odds (CEO) approach proposed by Pleiss et al.[71] While originally designed for binary classification, it was adapted to the multiclass cognitive impairment detection setting using a one-vs-rest framework.

The method combines group-wise probability calibration and threshold optimization to satisfy Equalized Odds across protected subgroups. For each diagnostic class $c \in Y$ and subgroup $a \in A$, calibrated probabilities are obtained as: $\hat{p}_{a,c} = Calib(f_\theta(x_{a,c}))$, where $Calib(.)$ denotes isotonic calibration fit on the validation set for subgroup $a$ in the one-vs-rest task for class $c$.

For each subgroup–class pair, an optimal classification threshold $t_{a,c}$ is identified by solving a constrained optimization problem that minimizes deviations in TPR and FPR between groups while preserving overall utility:

$$\min_{t_{a,c}} \sum_a \left( \left| \text{TPR}_{a,c} - \overline{\text{TPR}}_c \right| + \left| \text{FPR}_{a,c} - \overline{\text{FPR}}_c \right| \right),$$

*subject to $t_{a,c} \in [0,1]$, where $\overline{TPR}_c$ and $\overline{FPR}_c$ are class-wise averages across subgroups.*

This threshold optimization produced subgroup-specific threshold that approximately equalize TPR and FPR while maintaining calibrated decision probabilities. The optimization is repeated for each class in a one-vs-rest framework, yielding per-class decision probabilities. Final multiclass predictions are obtained by selecting the class with the highest decision probability.

## 2.7 External Generalizability Evaluation- DementiaBank Chou Corpus (Mandarin)

To assess how well our models generalize beyond the multilingual PREPARE dataset, we conducted external validation on an independent Mandarin-speaking cohort. This step was necessary because the PREPARE corpus contains a systematic language–label confound: all Mandarin recordings are labeled as MCI, with no Mandarin control samples. As a result, both models predicted the MCI class for every Mandarin input, preventing meaningful evaluation of language-specific performance and making it impossible to conduct any language-specific bias analysis for Mandarin.

To address this limitation, we used the Chou Corpus[72] from DementiaBank, which provides a more balanced Mandarin dataset suitable for evaluating cognitive-impairment detection. This corpus includes: (i) three picture description tasks—a father taking care of his baby, a night market, and park activities, (ii) recordings in Mandarin Chinese, and (iii) binary labels: MCI vs. control. The dataset includes 87 participants (47 MCI, 40 controls), with no information on age or education.

We performed a participant-level split, assigning ~60% of speakers to training (n = 51), ~20% to validation (n = 17) and the remaining ~20% (n = 19) to a held-out test set, ensuring all recordings from an individual appear in only one partition.

For transfer learning, both SpeechCARE-AGF and Whisper-LWF-LoRA were first fine-tuned on the PREPARE dataset, then further fine-tuned on the Chou training set using the same hyperparameters. Because the Chou dataset lacks demographic attributes, we report only F1-score and AUC as performance metrics, averaged across five random seeds with 95% confidence intervals computed across seeds.

## 3. Result

### 3.1 Performance evaluation

Both models demonstrated strong and stable performance across five random seeds. The SpeechCARE-AGF model achieved a mean F1-score of 70.87 ± 0.21, an AUC-ROC of 86.72 ± 0.92 (**Figure 7-a**) and an AUPRC (area under the precision–recall curve) of 75.04 ± 2.58 (**Figure 7-b**). This model received a special recognition award from the NIA PREPARE Challenge Phase 2. Across 67 competing teams, multi-class log-loss scores ranged from 1.5151 (worst) to 0.6299 (best). SpeechCARE-AGF achieved a log-loss of 0.6553, placing fourth and earning the special recognition award. The Whisper-LWF-LoRA model showed slightly higher performance, with an F1-score of 71.46 ± 0.34, an AUC-ROC of 87.58 ± 0.61 (**Figure 7-c**), and an AUPRC of 77.34 ± 1.16 (**Figure 7-d**). Notably, its precision–recall performance exhibited smaller across-seed variability than AGF, indicating more stable sensitivity–precision behavior across seeds. These results confirm that both architectures provide strong and stable baselines for multiclass cognitive impairment detection, suitable for subsequent analysis of bias and fairness mitigation techniques.

### 3.2 Fairness Evaluation

We evaluated model fairness across demographic and linguistic subgroups using Equality of Opportunity (EOO), which measures consistency in true positive rates, and Equalized Odds (EO), which captures disparities in both true positive and false positive rates. Our objective was not only to quantify subgroup disparities but also to analyze how each model architecture, SpeechCARE-AGF and Whisper-LWF-LoRA, responds to different bias-mitigation strategies. This allowed us to assess how architectural design influences the effectiveness of mitigation methods and to identify model-specific patterns in fairness behavior.

**Baseline Bias Evaluation.** To assess EOO, we compared TPR across subgroups, and to evaluate EO, we examined both TPR and FPR disparities (**Figure 8**). For both SpeechCARE-AGF and Whisper-LWF-LoRA, substantial disparities were observed across age and language. Age-related disparities were particularly pronounced: younger participants (ages 46–65) consistently showed higher TPRs, whereas the oldest group (≥80) had noticeably lower detection rates, indicating reduced model sensitivity for older adults despite their clinical importance. Mandarin-speaking participants exhibited extremely high TPRs (close to 1.0) for the MCI class, a direct reflection of the language–label confound in the PREPARE dataset where all Mandarin samples belonged to the MCI class; corresponding FPR estimates within Mandarin were not meaningful due to the absence of Mandarin control/AD samples. As a result, both models learned to assign the MCI label to nearly all Mandarin samples. Spanish speakers also showed lower TPRs compared to English speakers (44.54% vs. 57.01%) in the baseline setting, indicating reduced sensitivity in both models. We hypothesize that this disparity is primarily related to task distribution rather than inherent linguistic differences. Specifically, over 90% of the Spanish language data were collected through sentence reading tasks, whereas English data incorporated a more diverse set of elicitation methods, including picture description tasks. Sentence reading tasks may provide insufficient linguistic biomarkers for cognitive impairment detection compared to picture description task.

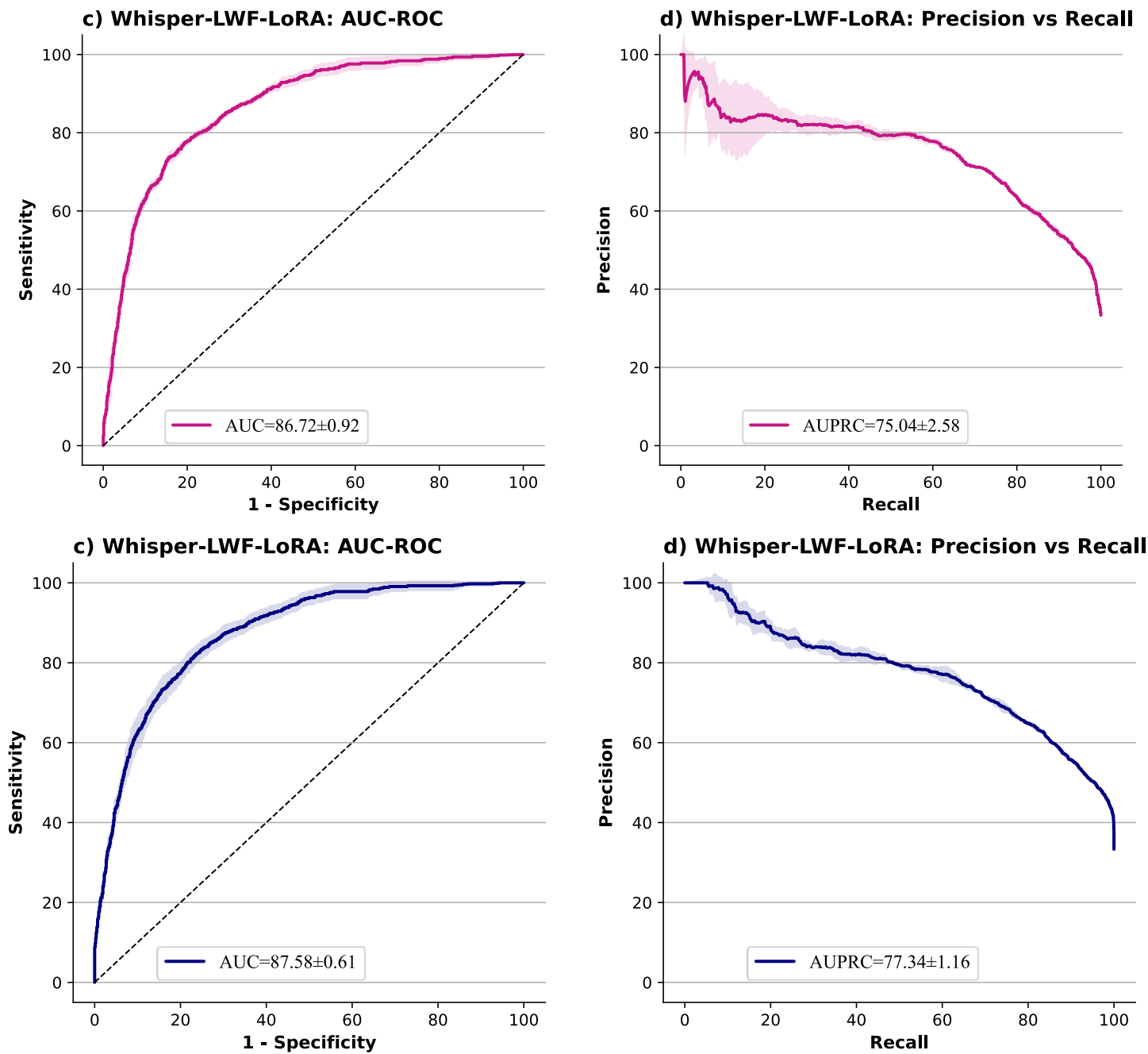

**Figure 7.** Model performance evaluated using AUC and precision–recall curves for SpeechCARE-AGF and Whisper-LWF-LoRA.

Across both models, education level also influenced fairness, with participants holding advanced/graduate degrees consistently achieving lower TPRs and higher FPRs. This may be due to the small number of participants with graduate degrees in the dataset. Gender differences were relatively small, with males showing slightly higher FPRs.

**Individual Bias Mitigation.** We applied all bias-mitigation techniques described in Section 2.6 to each demographic subgroup, with particular focus on age and language, which showed the largest baseline disparities in both architectures. Our evaluation covered all pre-processing approaches (generative and non-generative oversampling), in-processing and post-processing techniques. For clarity of presentation, we report only the best-performing techniques within each category (pre-processing, in-processing, and post-processing). We also present TPR-based comparisons in the main text because they directly correspond to EOO. For EO, which additionally requires examining FPRs, we include the full FPR plots and subgroup results in the **Appendix 1** to keep the main Results section concise.

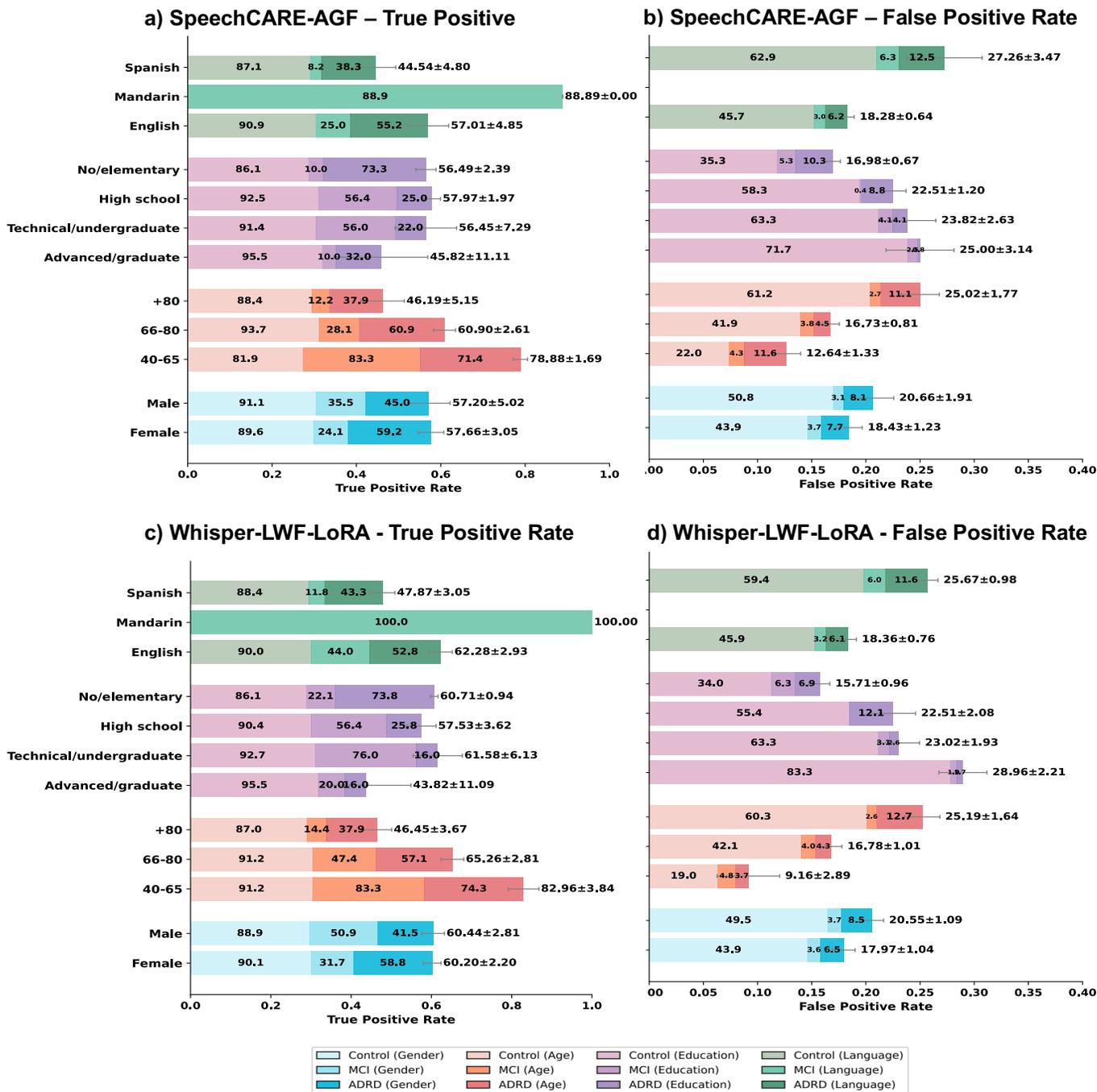

**Figure 8.** True Positive Rate (TPR) and False Positive Rate (FPR) disparities across demographic and linguistic subgroups for the SpeechCARE-AGF and Whisper-LWF-LoRA models before bias mitigation. (a) and (c) show subgroup-level TPRs for SpeechCARE-AGF and Whisper-LWF-LoRA, respectively, with variations across subgroups indicating violations of equality of opportunity. (b) and (d) present the corresponding FPRs. Together, disparities in both TPR and FPR indicate violations of equalized odds. Error bars reflect 95% confidence intervals over five random seeds.

**Figure 9** shows gender-specific TPRs for SpeechCARE-AGF and Whisper-LWF-LoRA before and after applying the best-performing mitigation techniques. Because TPR disparities directly reflect EOO, these results highlight how different mitigation strategies shift subgroup sensitivity. Although voice conversion yielded the strongest pre-processing performance relative to other oversampling approaches, its overall impact remained limited. For SpeechCARE-AGF, voice conversion produced only marginal improvements—most notably a small TPR increase for female speakers—while male performance showed minimal change. For Whisper-LWF-LoRA, voice conversion did not improve TPR for either gender, highlighting the model's limited compatibility with this oversampling technique.

In contrast, frequency reweighting provided the most consistent improvements in EOO across both models. For SpeechCARE-AGF, frequency reweighting increased female TPR (from 57.66% → 71.96%) and also improved male performance (from 57.20% → 62.07%). For Whisper-LWF-LoRA, this technique similarly boosted TPRs for both males (60.44% → 62.32%) and females (60.20% → 67.10%). Post-processing methods also produced moderate gains, where calibrated equalized-odds thresholds improved male TPR for Whisper-LWF-LoRA to 61.18%. To fully assess EO, which incorporates both TPR and FPR, corresponding FPR plots are reported in **Appendix 1**.

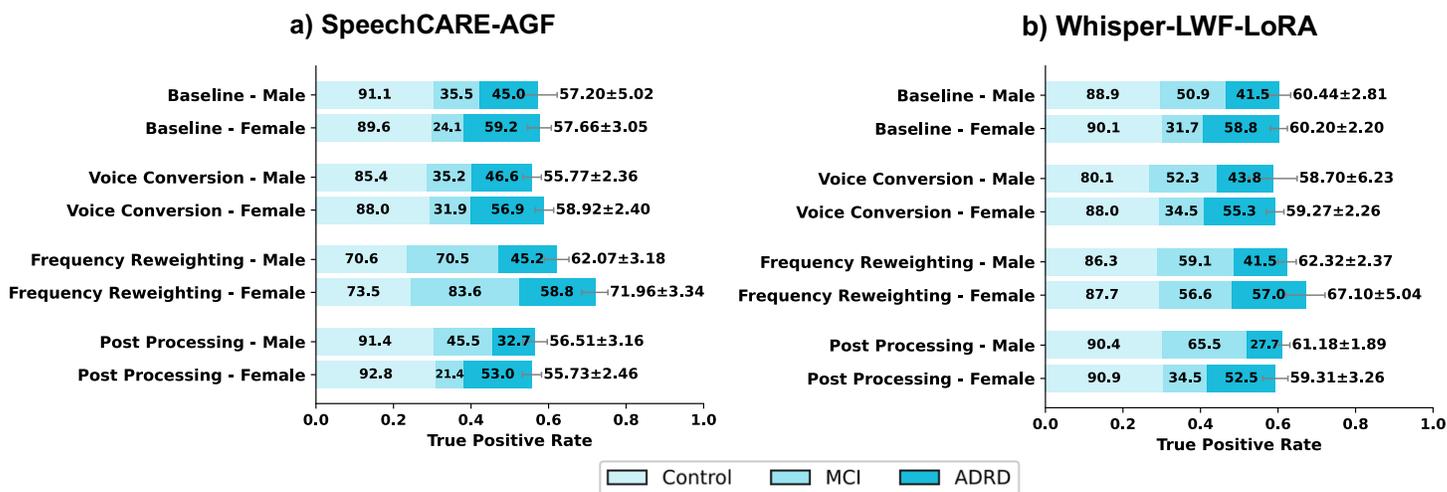

**Figure 9.** True positive rates (TPR) for male and female subgroups under baseline and bias-mitigation techniques for SpeechCARE-AGF and Whisper-LWF-LoRA. Disparities in TPR between genders indicate violations of equality of opportunity. Results highlight the best-performing methods from each category (pre-processing, in-processing, and post-processing), with error bars reflecting 95% confidence intervals over five random seeds.

**Figure 10** shows age-specific TPRs for SpeechCARE-AGF and Whisper-LWF-LoRA before and after applying the best-performing mitigation techniques. Full age-stratified results for all pre-processing, in-processing, and post-processing methods are reported in **Appendix 2**. Oversampling-based methods generally yield stronger improvements for SpeechCARE-AGF than for Whisper-LWF-LoRA, which may reflect architectural differences, including SpeechCARE-AGF's adaptive modality-fusion mechanism. The 80+ subgroup had the lowest baseline TPR, so oversampling methods were targeted to this group. Among pre-processing approaches, voice conversion gave the largest TPR gain for SpeechCARE-AGF (~49.9%), whereas time shifting was most effective for Whisper-LWF-LoRA (~50.4%). These oversampling strategies also substantially improved MCI detection in the 80+ group (SpeechCARE-AGF: 12.2% → 32.2%; Whisper-LWF-LoRA: 14.4% → 28.9%).

In-processing frequency reweighting produced the strongest overall reduction in age-related disparities, increasing 80+ TPRs to 59.38% for SpeechCARE-AGF and 55.14% for Whisper-LWF-LoRA. Post-processing Calibrated Equalized Odds (CEO) adjustments provided additional gains for Whisper-LWF-LoRA (80+ TPR: 59.28%) but at the cost of reduced TPR in younger adults (e.g., 46–65 TPR: 68.31%). These findings underscore that architectural design strongly shapes how models respond to different bias-mitigation strategies. Because Equalized Odds depends on both TPR and FPR, the corresponding age-specific FPR analyses are reported in **Appendix 1** to complement the TPR-based EOO findings shown here. Together, these results highlight that architectural design strongly influences how different mitigation strategies affect subgroup fairness.

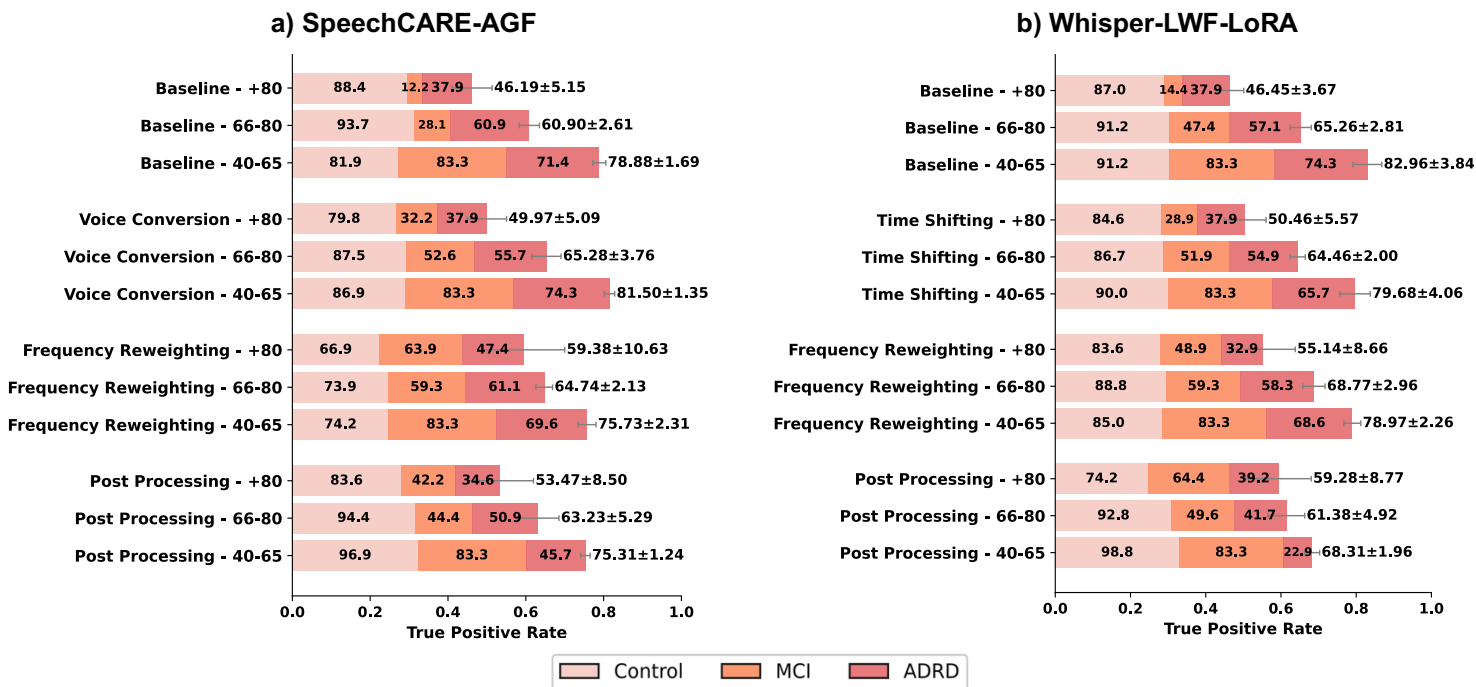

**Figure 10.** True positive rates (TPR) for age subgroups under baseline and bias-mitigation techniques for SpeechCARE-AGF and Whisper-LWF-LoRA. Results highlight the best-performing methods from each category (pre-processing, in-processing, and post-processing). Error bars reflect 95% confidence intervals over random seeds.

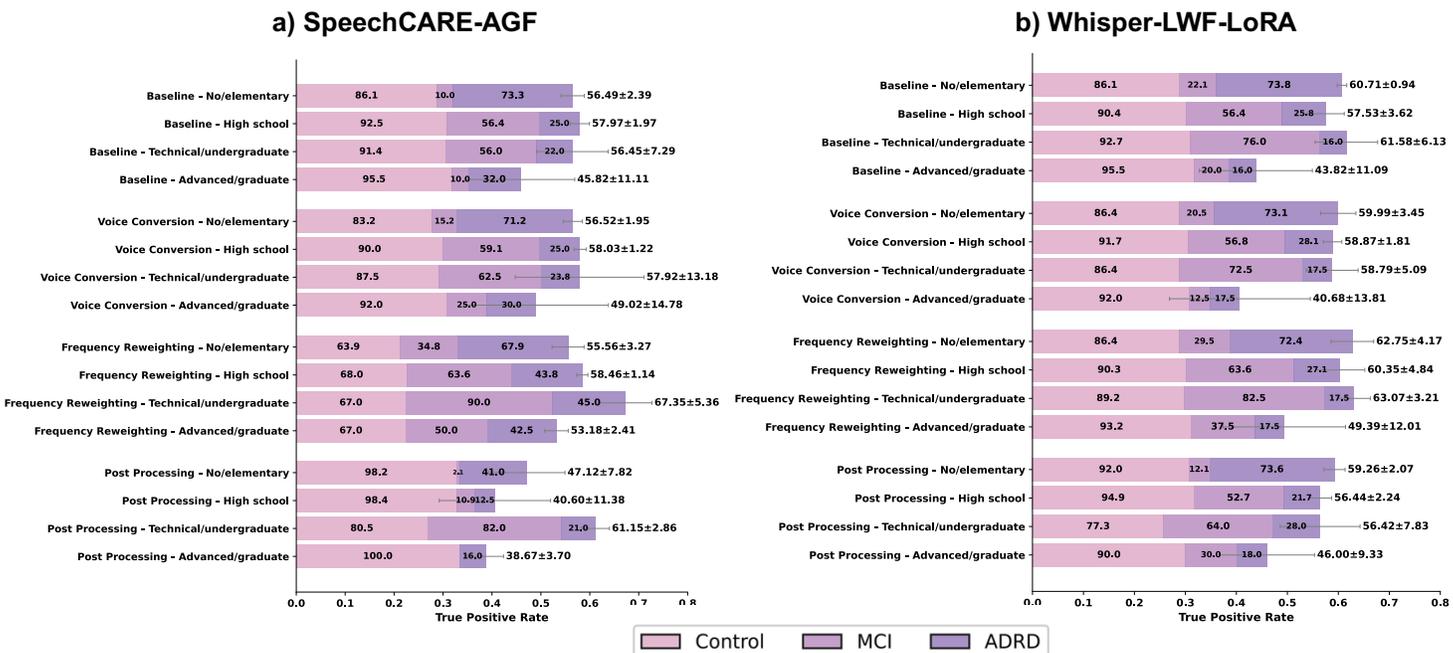

**Figure 11.** Effect of bias mitigation techniques across education subgroups for (a) SpeechCARE-AGF and (b) Whisper-LWF-LoRA. Each panel shows true positive rates (TPR) for Control, MCI, and AD classes across four education levels. Results compare the baseline models with three representative mitigation techniques: voice conversion (applied to the Advanced/Graduate subgroup), frequency reweighting, and post-processing (calibrated equalized odds).

**Figure 11** reports TPRs across education levels after applying the best-performing mitigation techniques. The Advanced/Graduate subgroup showed the lowest baseline TPR; therefore, oversampling was targeted to this subgroup. Among pre-processing approaches, voice conversion produced a modest improvement for SpeechCARE-AGF (TPR increased to 49.0%), driven largely by gains in the MCI class. In contrast, oversampling methods, including voice conversion did not benefit Whisper-LWF-LoRA and slightly reduced performance (43.82% → 40.68%).

In-processing frequency reweighting yielded the most consistent improvements across both architectures, increasing Advanced/Graduate TPRs to 53.18% for SpeechCARE-AGF and 49.39% for Whisper-LWF-LoRA, while preserving stable performance in the remaining education subgroups. Post-processing adjustments yielded only modest improvements for Whisper-LWF-LoRA and even reduced TPR for the Advanced/Graduate subgroup in SpeechCARE-AGF, showing that the two architectures respond differently to post-processing for education-related disparities.

**Figure 12** reports language-specific TPRs after applying the best-performing mitigation techniques, providing direct insight into EOO. Among pre-processing methods, none of the oversampling approaches, including voice conversion, which was previously helpful for demographic bias mitigation, improved results in this setting. In contrast, the language-specific encoder SpanBERTa produced the largest gains for Spanish speech in SpeechCARE-AGF, increasing TPR to 53.33%. For Whisper-LWF-LoRA, frequency reweighting yielded the strongest improvement, raising Spanish TPR to 53.58% and also benefiting English speakers (62.28% → 67.50%).

For Mandarin speakers, all mitigation techniques maintained their already high TPRs but, as expected, could not resolve the underlying label–language confound; therefore, Mandarin-specific performance and fairness conclusions remain limited in PREPARE dataset. Post-processing methods had no measurable effect on Spanish performance and only minimal effects on English and Mandarin across Whisper-LWF-LoRA. Because Equalized Odds requires evaluating both TPR and FPR, the corresponding language-specific FPR results are reported in **Appendix 1** to complement these TPR-based EOO findings.

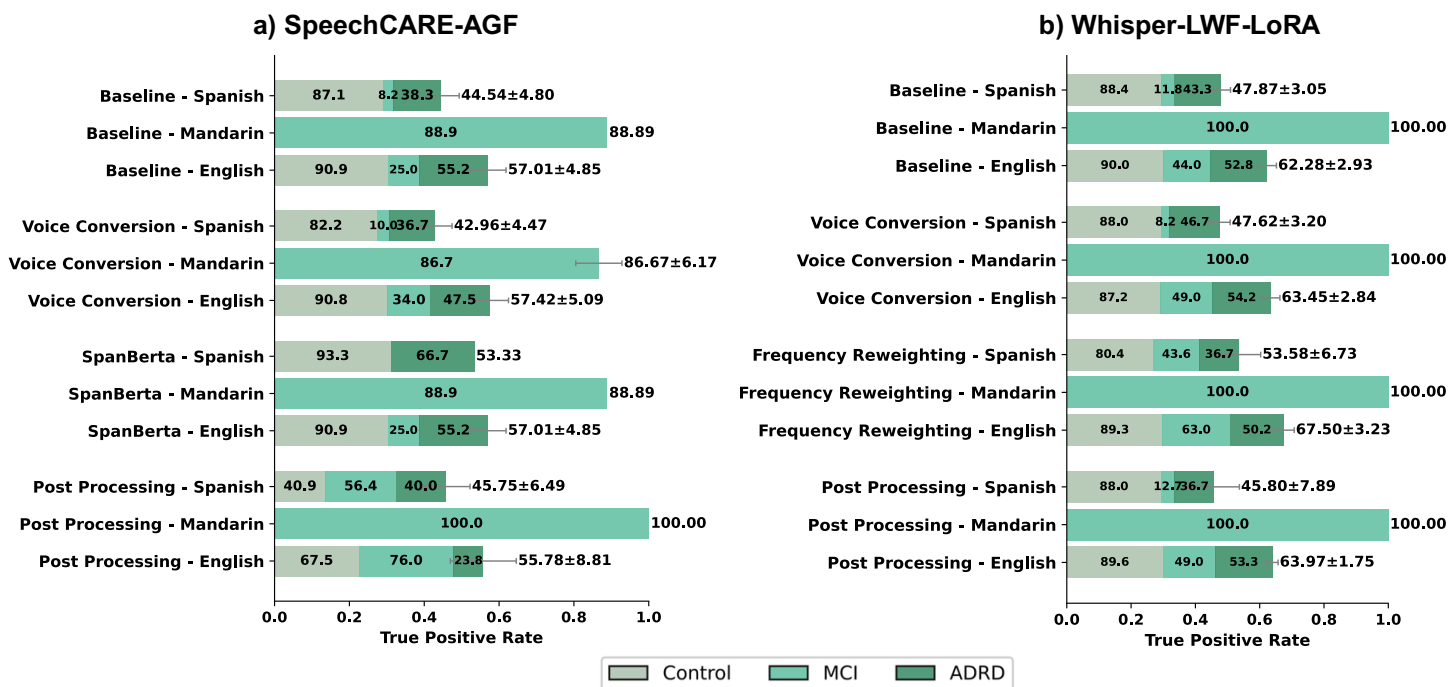

**Figure 12.** True Positive Rates (TPR) across language subgroups (Spanish, Mandarin, English) for (a) SpeechCARE-AGF and (b) Whisper-LWF-LoRA under different bias-mitigation techniques. Results include baseline performance and the best-performing pre-processing (voice conversion), in-processing (SpanBERTa, and frequency reweighting), and post-processing methods. SpanBERTa improves Spanish TPR for SpeechCARE-AGF, while frequency reweighting benefits Spanish and English TPRs for Whisper-LWF-LoRA. Mandarin TPRs remain uniformly high across all methods due to the language–label confound in the source data.

**Combined bias mitigation techniques.** To further reduce demographic disparities, we evaluated combinations of mitigation techniques spanning pre-processing, in-processing, and post-processing stages. For SpeechCARE-AGF, the most effective configuration combined voice-conversion oversampling targeted to the 80+ subgroup with age-based frequency reweighting. This pairing produced the largest and most consistent improvements in both TPR and FPR across subgroups, substantially enhancing both Equality of Opportunity and Equalized Odds relative to baseline. Figure 13-a,b presents subgroup-specific TPRs and FPRs for the SpeechCARE-AGF model under this combined mitigation approach.

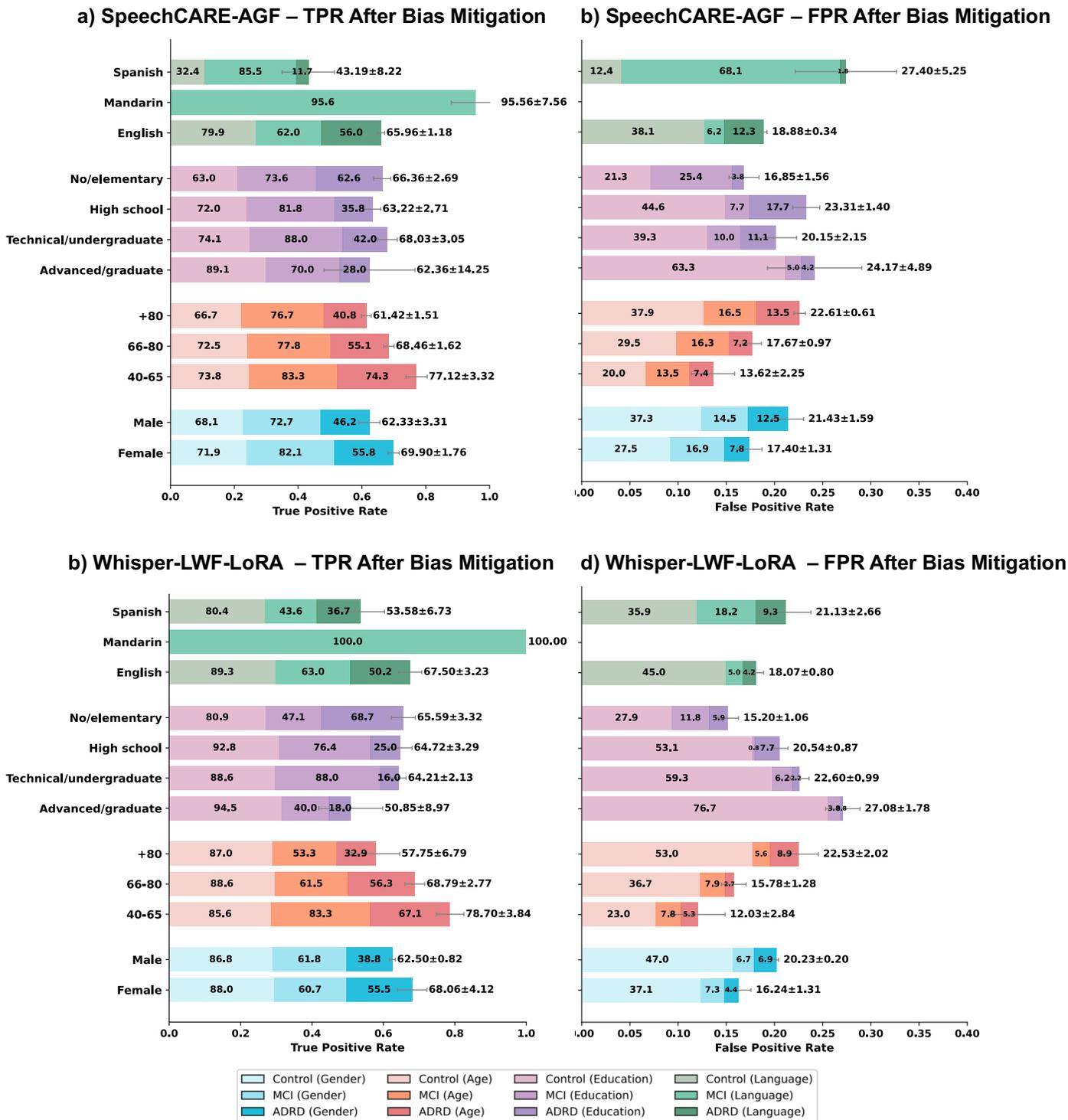

**Figure 13.** True Positive Rate (TPR) and False Positive Rate (FPR) across demographic subgroups after applying the best-performing bias-mitigation techniques for (a,b) SpeechCARE-AGF and (c,d) Whisper-LWF-LoRA. Error bars indicate 95% confidence intervals across five runs.

Following mitigation, the oldest age group (80+) showed marked improvement in TPR (46.19% → 61.42%), while the 40-65 and 66-80 groups maintained strong performance (77.12% and 68.46%, respectively). Education-related disparities also narrowed, with the Advanced/Graduate subgroup improving from 45.82% to 62.36% TPR, bringing it closer to other education levels (63.22%–68.03%). Both male and female participants showed substantial TPR increases (male: 57.20% → 62.33%; female: 57.66% → 69.90%), though gains were larger for female participants. Language-specific performance remained challenging, with Spanish performance changing only marginally (TPR: 43.19%, FPR: 27.40%), while English maintained strong TPR (65.96%). FPR patterns showed complementary improvements: the 80+ group achieved lower false positive rates (25.02% → 22.61%), and most education and gender subgroups showed reductions in FPR, reflecting improved Equalized Odds.

For Whisper-LWF-LoRA, we evaluated combinations of pre-, in-, and post-processing techniques. However, adding pre-processing methods (time shifting, voice conversion, TTS) to in- or post-processing strategies did not enhance performance and in some cases degraded results compared to in-processing alone. The most effective configuration was language-based frequency reweighting alone, which produced reliable improvements across demographic and linguistic subgroups (**Figure 13-c,d**), yielding the largest gains in both Equality of Opportunity and Equalized Odds relative to baseline.

Language-based frequency reweighting substantially improved age-related fairness, with the 80+ group achieving 57.75% TPR (up from 46.45% baseline) while preserving strong performance in younger groups (78.70% for 40-65, 68.79% for 66-80). Education-related performance became more balanced, with Advanced/Graduate TPR increasing to 50.85% and other education levels ranging from 64.21% to 65.59%. Gender performance improved for both groups (female: 60.20% → 68.06%; male: 60.44% → 62.50%), with female participants showing larger gains. Language-specific results showed notable progress for Spanish speakers (TPR: 47.87% → 53.58%) and continued strong English performance (67.50%). FPR reductions were observed across most subgroups, particularly for age (80+: 25.19% → 22.53%) and language (Spanish: 25.67% → 21.13%), demonstrating improved specificity alongside sensitivity gains and better overall Equalized Odds.

## 3.4 External Evaluation of SpeechCARE-AGF and Whisper-LWF-LoRA on Mandarin Speech for MCI Detection

To assess how well our models generalize beyond the multilingual PREPARE dataset, we conducted external validation on an independent Mandarin-speaking cohort. This step was necessary because the PREPARE corpus contains a systematic language–label confound: all Mandarin recordings are labeled as MCI, with no Mandarin control samples. As a result, both models predicted the MCI class for every Mandarin input, preventing meaningful evaluation of language-specific performance and making it impossible to conduct any language-specific bias analysis for Mandarin.

Using the transfer learning approach, fine-tuning SpeechCARE-AGF and Whisper-LWF-LoRA on the Mandarin-speaking Chou dataset resulted in an F1-score of 83.10 ± 11.13% and 83.56 ± 10.04 respectively, with 95% confidence intervals calculated across 5 runs. These results demonstrate that both models are capable of adapting to Mandarin speech when fine-tuned on representative Mandarin training data. This finding highlights the critical role of linguistic diversity in model development to ensure robust and equitable performance across languages in cross-lingual cognitive impairment screening applications.

## 3.5 Modality Weight Analysis for Oversampling Method (Voice Conversion) and Adversarial Debiasing

**Voice conversion.** Across our experiments, oversampling strategies targeting the 80+ subgroup (e.g., +80 voice-conversion oversampling) produced consistently larger gains for SpeechCARE-AGF than for Whisper-LWF-LoRA (**Appendix 2**). This difference reflects fundamental architectural properties: the two models vary in how effectively they can incorporate additional samples that increase within-class acoustic and prosodic variability, a characteristic typical of older-adult speech (e.g., greater variability in acoustic characteristics such as pitch, rhythm, and articulation).

To investigate this disparity, we examined modality weighting patterns in both models. As shown in **Figure 14**, SpeechCARE-AGF's participant-specific gating network adaptively reweighted modalities after +80 voice-conversion oversampling, shifting from balanced weights (52.73% linguistic, 47.27% acoustic) to linguistic-dominant patterns (76.58% linguistic), effectively prioritizing linguistic features while reducing reliance on the acoustic characteristics. In contrast, Whisper-LWF-LoRA's global fusion coefficients remained stable (≈62% acoustic, ≈38% linguistic before and after), reflecting its limited capacity to rebalance modalities in response to augmented data. This stability is partly attributable to Whisper's inherent cross-attention mechanism between encoder and decoder, which creates tight coupling between acoustic and linguistic representations that resists dynamic reweighting.

**Adversarial debiasing** produced a contrasting reweighting pattern. For SpeechCARE-AGF, adversarial training reversed the modality preference toward acoustic features (54.07% acoustic, 45.93% linguistic), likely because linguistic features correlate more strongly with protected attributes (e.g. age, language) than low-level acoustic cues. The gating network thus adaptively down weighted the linguistic branch to satisfy fairness constraints while preserving diagnostic utility through acoustic information. Whisper-LWF-LoRA's fusion weights again remained unchanged (≈62% acoustic, ≈38% linguistic), with the encoder-decoder cross-attention architecture preventing adaptive modality rebalancing in response to the adversarial objective.

These findings reveal that fairness intervention effectiveness depends critically on architectural flexibility. SpeechCARE-AGF's participant-level adaptive gating enables flexible responses to both data-level (oversampling) and optimization-level (adversarial) interventions through dynamic modality reallocation. Whisper-LWF-LoRA's fixed fusion structure and cross-attention coupling limit adaptability. This architectural difference is a critical determinant of which fairness strategies succeed in practice.

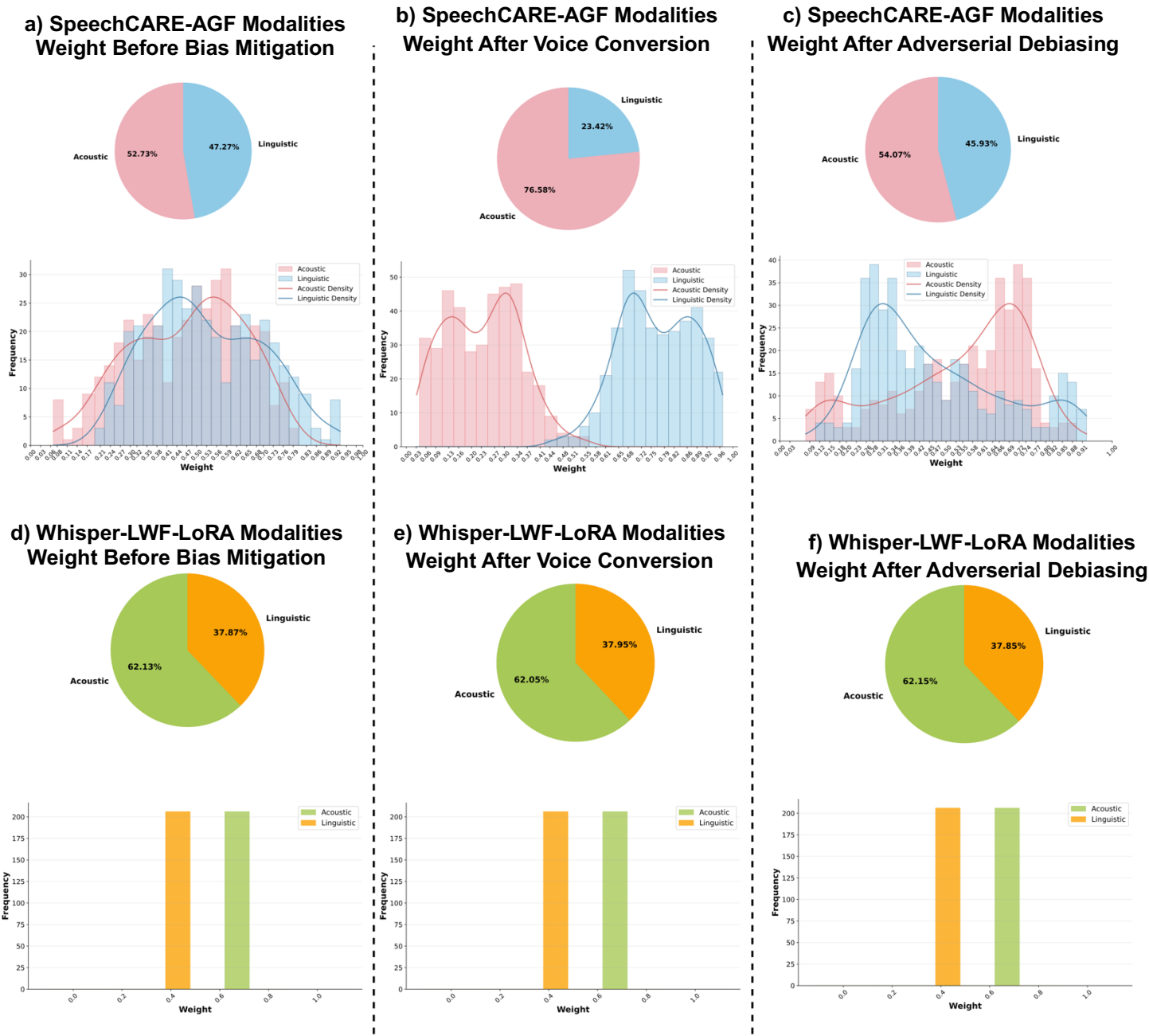

**Figure 14.** Modality Weight Analysis for SpeechCARE-AGF and Whisper-LWF-LoRA Before and After Bias Mitigation. The top panels show the distribution of adaptive modality weights (linguistic vs. acoustic) produced by the SpeechCARE-AGF gating network. Oversampling shifts AGF's modality weighting toward greater reliance on linguistic features for most participants. The bottom panels show modality weights for Whisper-LWF-LoRA, whose encoder–decoder fusion weights remain stable before and after bias mitigation.

# 4. Discussion

This study provides the first systematic evaluation of fairness-oriented mitigation techniques for speech-based detection of cognitive impairment across pre-processing, in-processing, and post-processing stages. Using two distinct architectures, SpeechCARE-AGF and Whisper-LWF-LoRA, and a large multilingual clinical dataset, we examined subgroup performance across gender, age, education, and language and assessed how different mitigation techniques influence equity in model predictions. Our findings demonstrate that fairness interventions are not uniformly effective across model architectures or demographic attributes; rather, their effectiveness depends critically on the representation learning mechanisms and fusion strategies embedded within each model.

Bias in transformer-based speech systems can arise from both transformer architecture and pretraining data characteristics, and our baseline results show how these factors differentially affect subgroup performance. Assessing equality of opportunity through TPR comparisons across subgroups revealed systematic disparities linked to model architecture and pretraining paradigms. Whisper-LWF-LoRA, built on Whisper's 680,000 hours of supervised speech–text pairs, learns tightly coupled acoustic–linguistic representations that generalize more evenly across speakers. Its relatively balanced baseline TPRs for gender (≈60% for males and females) and education levels demonstrate equality of opportunity across these attributes. In contrast, SpeechCARE-AGF relies on mHuBERT, whose self-supervised pretraining emphasizes low-level acoustic patterns without linguistic supervision, making the model more sensitive to age-related acoustic variability—such as reduced articulation rate and greater prosodic irregularity—which aligns with the markedly lower TPR observed for adults aged 80+ (~46%), representing a substantial violation of equality of opportunity across age groups. For Spanish speakers, Whisper-LWF-LoRA also showed stronger baseline performance, likely because its supervised encoder–decoder structure captures cross-lingual phonetic and prosodic patterns more effectively than the independently pretrained mHuBERT–mGTE components in SpeechCARE-AGF. Mandarin TPRs were artificially inflated due to the language–label confound, where all Mandarin samples were labeled MCI. Overall, these findings indicate that large-scale supervised multilingual pretraining supports more stable subgroup performance and better equality of opportunity, whereas models incorporating unsupervised acoustic encoders are more vulnerable to demographic variability, particularly age-related shifts in speech production.

Model architecture is another key factor influencing bias and the effectiveness of mitigation techniques. SpeechCARE-AGF uses a dynamic gating fusion mechanism that can reweight acoustic and linguistic modalities in response to new training distributions. Under voice-conversion oversampling, SpeechCARE-AGF shifted from a balanced weighting (~53% linguistic, 47% acoustic) to a linguistic-dominant pattern (~77%), reducing the impact of acoustic noise and age-related variability introduced by synthetic speech. This adaptability explains why voice conversion improved SpeechCARE-AGF's baseline fairness for older adults. In contrast, Whisper-LWF-LoRA uses a fixed encoder–decoder layer-weighted fusion, where modality contributions remain stable (~62% acoustic, ~38% linguistic) even after oversampling. The tight coupling between encoder and decoder representations, combined with subject-invariant fusion weights, constrains the model's ability to selectively reweight modalities in response to augmented data or individual-level variability, resulting in limited gains from voice conversion and, in some cases, performance degradation.

Speech-based screening algorithms naturally encode demographic information, particularly gender and age, because these cues are strongly embedded in acoustic features such as pitch and prosody. When models learn these cues as predictive shortcuts, they may inadvertently link them to diagnostic labels, creating subgroup disparities. In-processing techniques aim to counter this by altering the training objective so the model reduces reliance on demographic signals. In our experiments, adversarial debiasing—designed to penalize representations that retain demographic information—shifted internal features but yielded limited fairness improvements, with major subgroup gaps remaining close to baseline. Frequency reweighting, which increases the loss contribution of underrepresented subgroup–label combinations, produced more consistent gains across both architectures by improving equality of opportunity across age and language.

The source and mechanism of speech data collection can also impose substantial bias, particularly when certain groups are underrepresented or when the elicitation method differs across subgroups. These biases extend beyond simple sample imbalance and reflect the structure of the speech corpus itself. For example, Spanish speakers showed lower TPRs in both models because their recordings consisted primarily of sentence-reading tasks, which provide fewer semantic and discourse cues relevant to cognitive-impairment detection than picture-description tasks used for other subgroups. Conversely, Mandarin speakers exhibited artificially inflated TPRs due to a language–label confound, in which all Mandarin samples were labeled MCI, causing the models to learn an incorrect deterministic association. Additional factors, such as subgroup-specific preprocessing or recording-environment differences can further distort performance. These considerations underscore the need for research teams to carefully examine speech datasets for structural flaws that may introduce systematic bias.

Speech-processing algorithms can function as standalone tools or be integrated with other data sources to enhance early detection of cognitive impairment. While the FDA's 2025 510(k) clearance of the Lumipulse G pTau217/β-Amyloid 1-42 Plasma Ratio blood test represents an important advance in minimally invasive biomarker detection, biological assays do not directly capture changes in everyday communication. Speech and language alterations, such as reduced fluency, disorganized syntax, or vocal pattern shifts often appear early and reflect functional decline that biomarkers cannot assess. Speech-based models, including those evaluated in this study address this gap by identifying subtle communication changes from short recordings, offering a behavioral dimension that complements biological measures and can further enhance diagnostic performance when combined with other modalities.

A key takeaway from this work is that there is no universal solution to algorithmic fairness in speech-based cognitive screening. The effectiveness of each mitigation technique depended on the interaction between model architecture, demographic subgroup, and dataset characteristics: voice conversion improved fairness for SpeechCARE-AGF but not Whisper-LWF-LoRA; frequency reweighting reduced age-related disparities but offered limited benefits for gender; and a Spanish RoBERTa model (SpanBERTa) improved Spanish performance but could not be integrated into Whisper's architecture in our current framework. These findings indicate that fairness techniques must be tailored to both the dataset and the model's inductive biases. Rather than relying on a single mitigation technique, researchers should evaluate combinations of pre-processing, in-processing, and post-processing techniques while accounting for data-collection procedures, speech tasks, and recording environments. Such architecture- and subgroup-specific techniques are essential for equitable clinical deployment of speech-based screening tools.

This study has several limitations. First, the PREPARE dataset contains structural imbalances—such as task-type differences for Spanish speakers and a language–label confound for Mandarin—that constrain evaluation of language-related fairness. Second, subgroup sample sizes, particularly for Mandarin and higher education levels, limit the precision of fairness estimates. Third, our analysis focused on two transformer-based architectures; findings may not generalize to other model families or clinical settings. Finally, while we evaluated multiple mitigation strategies, their effectiveness may vary under different recording conditions or speech tasks, underscoring the need for further validation in diverse real-world environments.

## Data Availability

The data used in this research is from the 2024 NIA PREPARE challenge, provided exclusively to our team as participants in the challenge. This dataset is not publicly available, and participants are prohibited from sharing it. To request access, please contact the challenge organizers [here]. Additionally, the data is now part of DementiaBank; access may also be requested by contacting the DementiaBank administrators directly.

## Acknowledgements

The authors have no acknowledgements to declare.

## Author Contributions

Yasaman Haghbin: Contributed to model development, model evaluation, and manuscript drafting.
Sina Rashidi: Contributed to model development, and manuscript drafting.
Ali Zolnouri: Contributed to model evaluation.
Maryam Zolnoori: Contributed to manuscript drafting.

## Competing Interests

Authors have no conflict of interest to declare.

## Human Ethics and Consent to Participate declarations
not applicable.

## Funding Declaration

# Appendix 1 – False Positive Rate (FPR) Evaluation Across Mitigation Techniques.

As shown in **Figure 15**, false positive rate (FPR) patterns often differ from TPR trends. For gender, voice conversion, frequency reweighting, and post-processing produce only modest shifts in FPR, with changes typically within ±3 percentage points for both models. Age-related FPR disparities show greater variability: oversampling the 80+ group reduces FPR for older adults in SpeechCARE-AGF, whereas Whisper-LWF-LoRA tends to exhibit smaller or inconsistent improvements. However, frequency reweighting reduces FPR in both models. For education subgroups, frequency reweighting generally provides the most stable FPR behavior, limiting the inflation observed with oversampling-based approaches. Language subgroups show the smallest FPR changes overall, reflecting the dominance of TPR-driven disparities, particularly for the Mandarin subgroup. Collectively, these FPR results illustrate how different mitigation strategies can introduce sensitivity–specificity trade-offs and complement the TPR analyses reported in the main text.

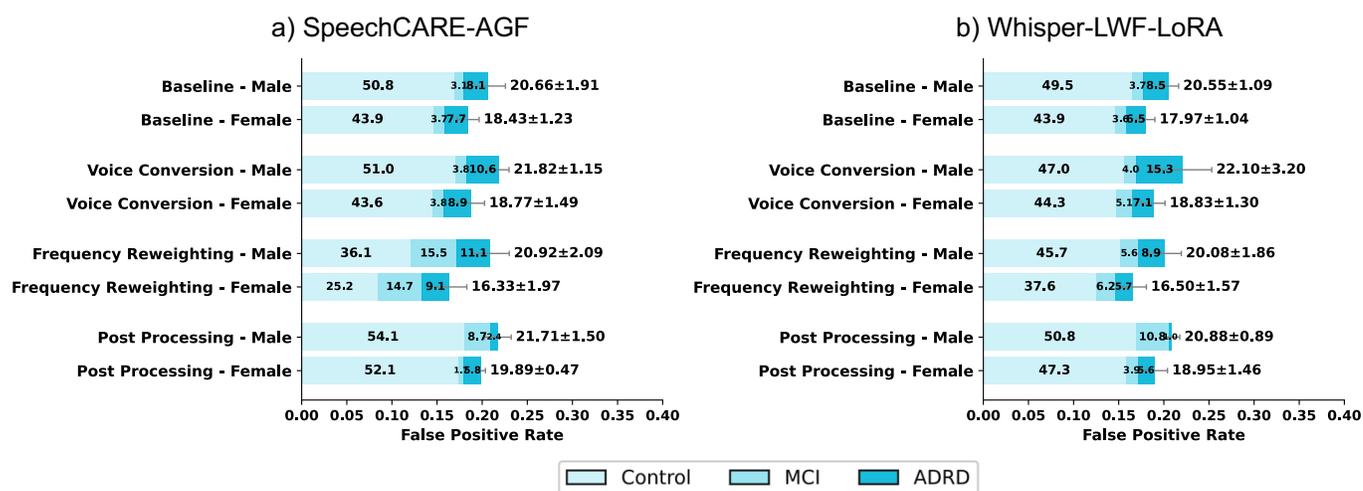

**Figure 15.** False positive rates (FPR) for male and female subgroups under baseline and bias-mitigation strategies for SpeechCARE-AGF and Whisper-LWF-LoRA. Results highlight the best-performing methods from each category. Error bars reflect 95% confidence intervals over five random seeds.

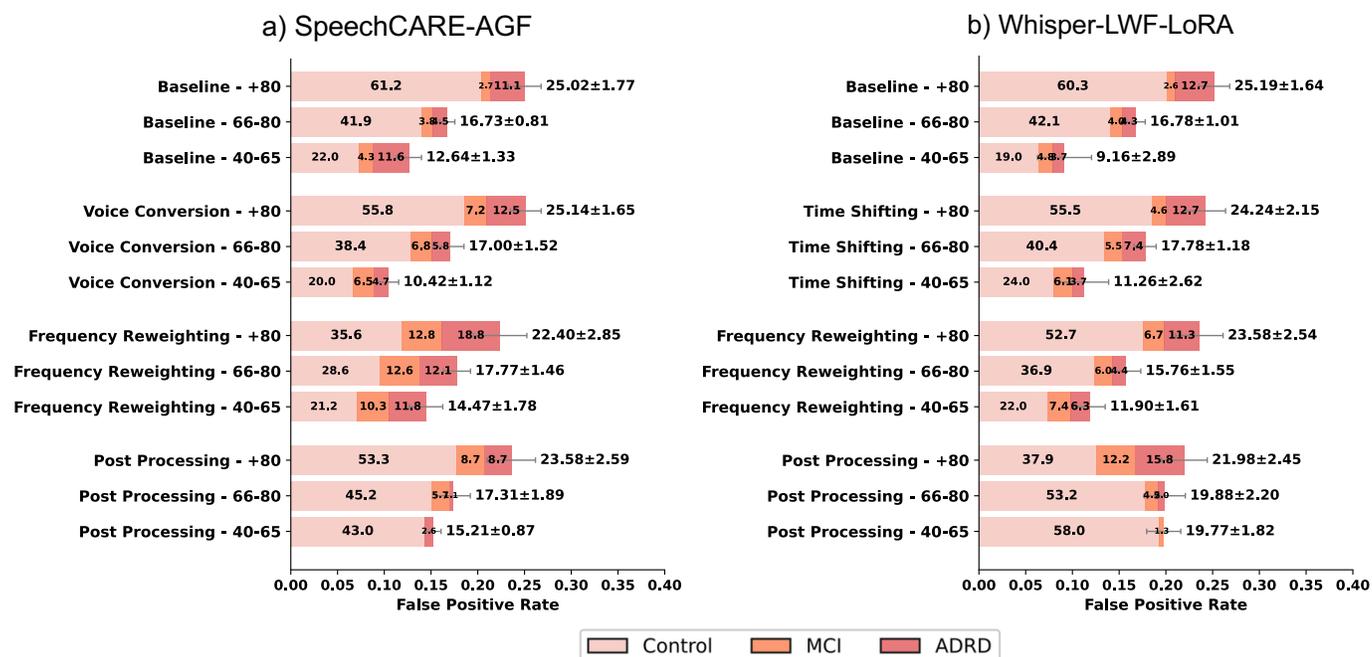

**Figure 16.** False positive rates (FPR) for age subgroups under baseline and bias-mitigation strategies for SpeechCARE-AGF and Whisper-LWF-LoRA. Results highlight the best-performing methods from each category with oversampling shown for the +80 subgroup due to its largest performance gap in baseline.

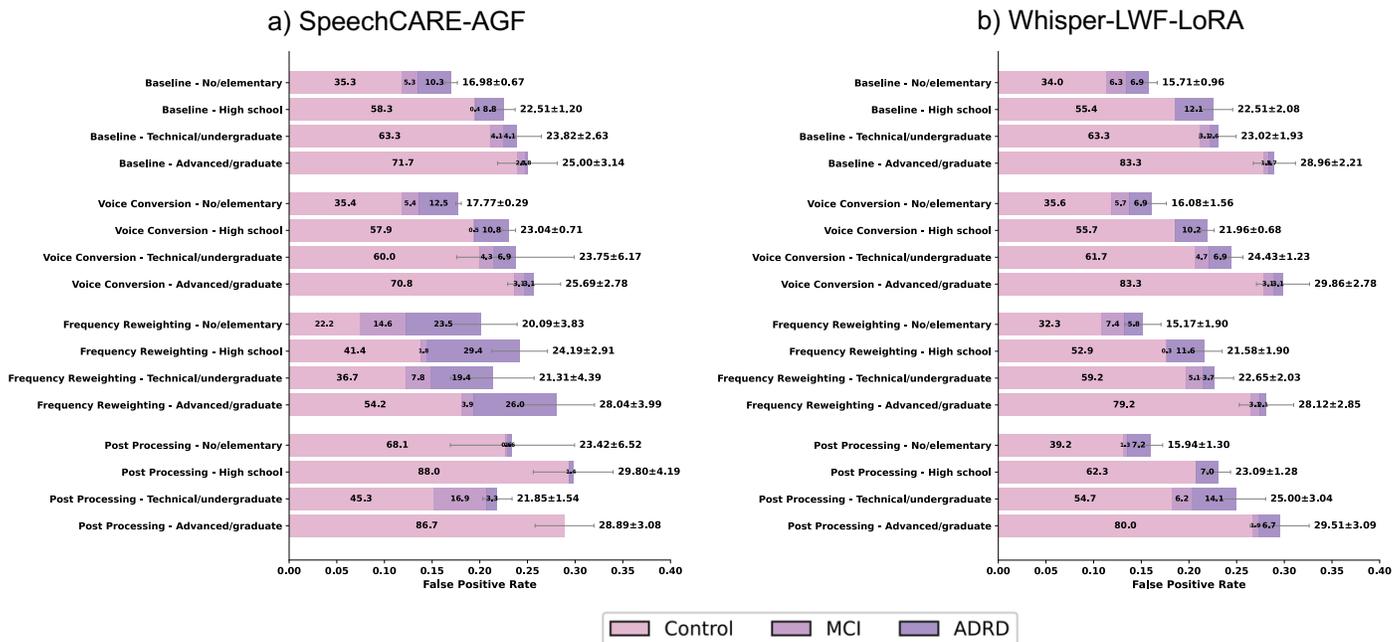

**Figure 17.** Effect of bias mitigation strategies across education subgroups for (a) SpeechCARE-AGF and (b) Whisper-LWF-LoRA. Each panel shows false positive rates (FPR) for Control, MCI, and AD classes across four education levels. Results compare the baseline models with three representative mitigation techniques: voice conversion (applied to the Advanced/Graduate subgroup), frequency reweighting, and post-processing (calibrated equalized odds).

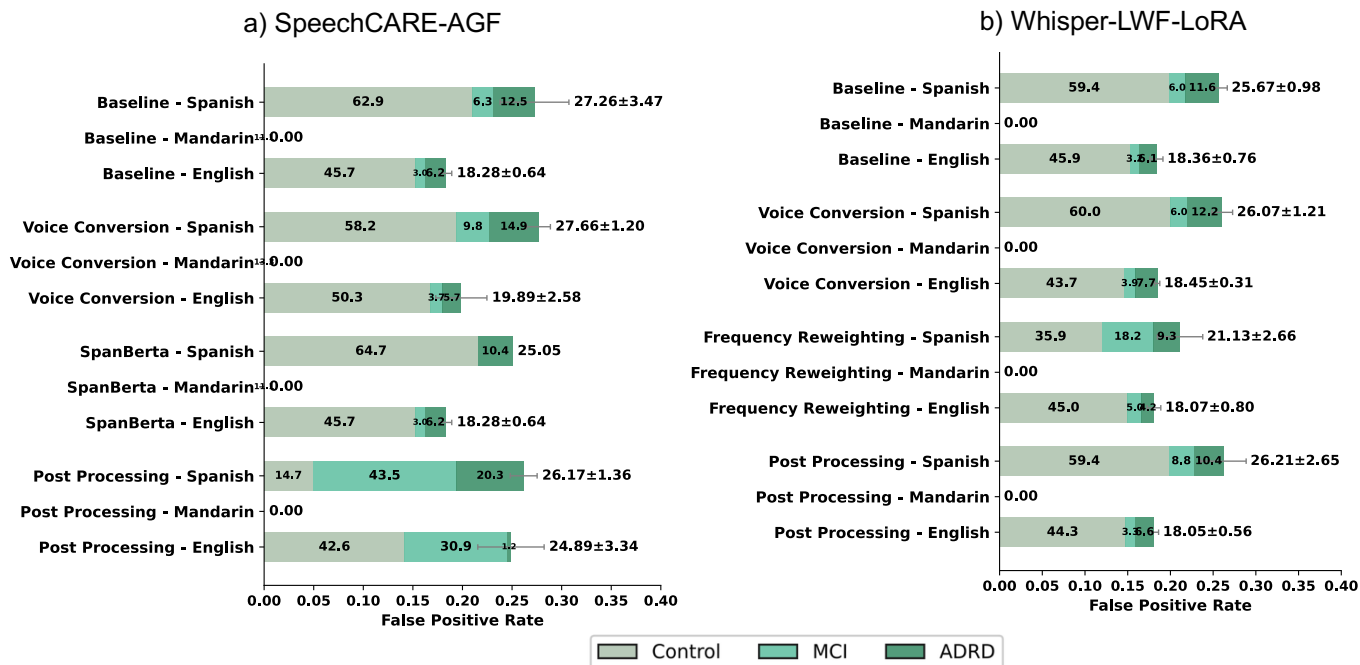

**Figure 18.** False Positive Rates (FPR) across language subgroups for SpeechCARE-AGF (a) and Whisper-LWF-LoRA (b) under different bias-mitigation strategies. Results include baseline performance and the best-performing pre-processing (Spanish voice conversion), in-processing (SpanBERTa, and frequency reweighting), and post-processing methods. Error bars reflect 95% confidence intervals over five random seeds.

# Appendix 2. The effect of all bias-mitigation methods on age-related subgroups for both SpeechCARE-AGF and Whisper-LWF-LoRA.

**Figure 19** summarizes the effect of all bias-mitigation methods on age-related subgroups (40–65, 66–80, 80+) for both (a) SpeechCARE-AGF and (b) Whisper-LWF-LoRA. For each model, the figure reports class-wise TPRs after applying every pre-processing (non-generative, and generative), in-processing (frequency and TPR reweighting, adversarial debiasing), and post-processing method (Calibrated Equalized odds). The panels illustrate how different strategies primarily boost sensitivity for the oldest (80+) group, often at some cost to the younger groups, and highlight that SpeechCARE-AGF generally achieves larger TPR gains for older adults in oversampling methods, whereas Whisper shows more modest or method-dependent improvements.

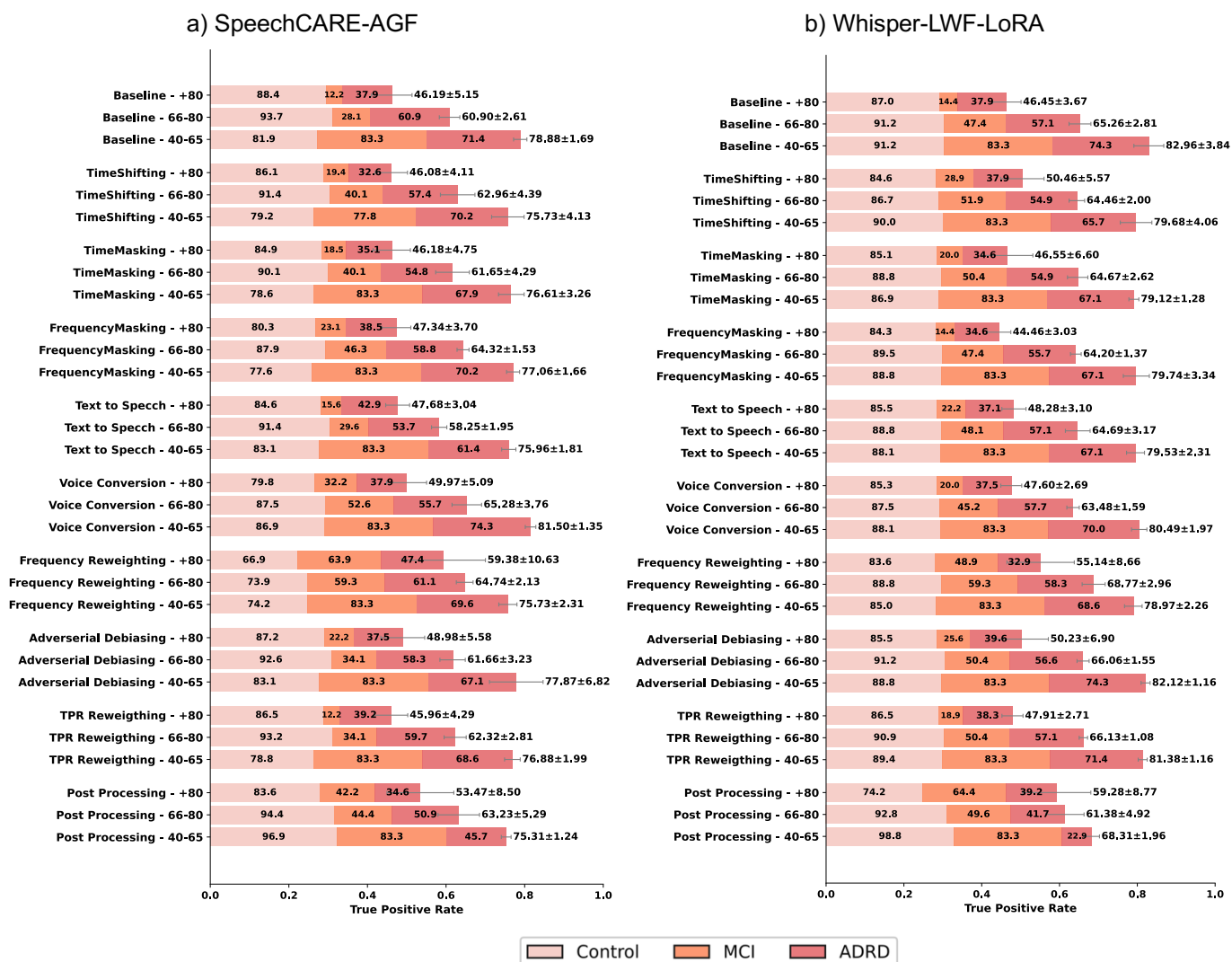

**Figure 19.** True Positive Rates (TPRs) for all bias-mitigation methods applied to age-related subgroups (40–65, 66–80, 80+) for (a) SpeechCARE-AGF and (b) Whisper-LWF-LoRA. The figure compares the impact of every pre-processing, in-processing, and post-processing technique, illustrating how different strategies affect sensitivity across age groups, with the largest improvements generally observed in the 80+ subgroup.